\documentclass[11pt, showpacs]{revtex4}
\usepackage{graphicx, epsfig}

\def\beq{\begin{equation}}
\def\eeq{\end{equation}}
\def\beqa{\begin{eqnarray}}
\def\eeqa{\end{eqnarray}}
\def\iar{\begin{array}{l}}
\def\ear{\end{array}}

\begin{document}

\title{Gauge dependence of on-shell and pole mass renormalization prescriptions}
\author{Yong Zhou}
\affiliation{Beijing University of Posts and Telecommunications, school of science P.O. Box 123, Beijing 100876, China}

\begin{abstract}
We discuss the gauge dependence of physical parameter's definitions under the on-shell and pole mass renormalization prescriptions. By two-loop-level calculations we prove for the first time that the on-shell mass renormalization prescription makes physical result gauge dependent. On the other hand, such gauge dependence doesn't appear in the result of the pole mass renormalization prescription. Our calculation also implies the difference of the physical results between the two mass renormalization prescriptions cannot be neglected at two-loop level.
\end{abstract}

\maketitle

\section{Introduction}

The conventional on-shell mass renormalization prescription has been present for a long time. It renormalizes the real part of particle's inverse propagator to zero at physical mass point. For boson the on-shell mass renormalization condition is \cite{c1,cin0}
\beq
  m^2-m_0^2+Re\Sigma(m^2)\,=\,0\,,
\eeq
where $m_0$ is the bare mass and $\Sigma$ is the boson's diagonal self energy (for vector boson it is the transverse diagonal self energy).  But recently people proposed a new mass renormalization prescription which renormalizes both the real and the imaginary parts of the particle's inverse propagator to zero at the (complex) pole of the particle's propagator, i.e. \cite{c2,c3}
\beq
  \bar{s}-m_0^2+\Sigma(\bar{s})\,=\,0\,,
\eeq
where $\bar{s}$ is the pole of the particle's propagator. Written $\bar{s}=m_2^2-i m_2\Gamma_2$, $m_2$ is defined as the physical mass of the particle \cite{c2}. Putting the expression of $\bar{s}$ into Eq.(2) one has \cite{c2,c3}
\beq
  m_2^2-m_0^2+Re\,\Sigma(\bar{s})\,=\,0\,, \hspace{8mm}
  m_2\,\Gamma_2\,=\,Im\,\Sigma(\bar{s})\,.
\eeq
By expanding Eqs.(3) at $\bar{s}=m_2^2$ one readily has (see Eq.(1)) \cite{c2,c3}
\beq
  m-m_2\,=\,\Gamma_2\,Im\,\Sigma^{\prime}(m_2^2)/2+O(g^6)\,,
\eeq
where $\Sigma^{\prime}(m_2^2)=\partial\Sigma(m_2^2)/\partial p^2$ and $g$ is a generic coupling constant. For unstable boson the r.h.s. of Eq.(4) is gauge dependent \cite{c2,c3}. So A. Sirlin et al. claim that the on-shell mass definition $m$ of unstable particles is gauge dependent, since the pole mass definition $m_2$ is gauge independent \cite{c2,c3,c4,c5}.

But the conclusion that the pole mass definition $m_2$ is gauge
independent has been proposed for not very long time. We still
need to search new and stricter proofs to prove this conclusion.
In this paper we will discuss if the pole mass definition is gauge
independent and investigate the difference of physical result
between the on-shell and pole mass renormalization prescriptions.
The arrangement of this paper is as follows: firstly we discuss
the gauge dependencies of the counterterms of gauge boson W and
Z's mass and the sine of the weak mixing angle under the on-shell
and pole mass renormalization prescriptions; then we discuss the
gauge dependence of the two-loop-level cross section of the
physical process $\mu\rightarrow\nu_{\mu}e^{-}\bar{\nu}_e$ under
the two mass renormalization prescriptions; Lastly we give the
conclusion.

\section{Gauge dependencies of physical parameter's counterterms under the on-shell and pole mass renormalization prescriptions}

The gauge invariance of Lagrangian always requires the bare physical parameters are gauge independent. The natural deduction of this conclusion is the counterterms of physical parameters should also be gauge independent \cite{c6}, since the bare physical parameter can be divided into physical parameter and the corresponding counterterm, and the physical parameter is of course gauge independent. This criterion could be used to judge which mass renormalization prescription is reasonable, in other words which mass definition is gauge independent. In the following we will discuss the gauge dependence of the counterterms of gauge boson W and Z's mass and the sine of the weak mixing angle under the on-shell and pole mass renormalization prescriptions. For convenience we only discuss the dependence of W gauge parameter $\xi_W$ in the $R_{\xi}$ gauge, and we only introduce physical parameter's counterterms (i.e. we don't introduce field renormalization constants). The computer program packages {\em FeynArts} and {\em FeynCalc} \cite{c7} have been used in the following calculations. Here we note there are some early two-loop-level calculations about the massive gauge boson's self energies in Ref.\cite{cin}.

From Eqs.(1,3) one has for massive gauge boson \cite{cin1}
\beqa
  \delta m^2&=&Re\Sigma^T(m^2)\,, \nonumber \\
  \delta m_2^2&=&Re\Sigma^T(m_2^2)+m_2\Gamma_2\,Im\,\Sigma^{T\prime}(m_2^2)+O(g^6)\,,
\eeqa
where $\Sigma^T$ denotes the transverse self energy of the gauge boson. The one-loop-level mass counterterms of W and Z have been proven gauge independent \cite{c5}. So we only need to discuss the two-loop-level case. Firstly $m$ and $m_2$ should be regarded as equal quantities, since both of them are regarded as the physical mass of the same particle. Therefore we find the two-loop-level difference of the two mass counterterm is $m\Gamma Im\Sigma^{T\prime}(m^2)$. Every part of this term contains gauge-parameter-dependent Heaviside functions (which come from the one-loop-level $Im\Sigma^{T\prime}$ \cite{c2,c3}). So in order to discuss the difference of the gauge dependence of the two mass counterterms we only need to calculate the gauge dependence of the singularities of the two-loop-level $Re\Sigma^T(m^2)$, because only the singularities of $Re\Sigma^T(m^2)$ in $Re\Sigma^T(m^2)$ contain Heaviside functions. In other words for our purpose we only need to discuss the gauge dependence of the part which contains Heaviside functions of the two mass counterterms.

The two-loop-level self energies can be classified into two kinds: one kind contains one-loop-level counterterms, the other kind doesn't contain any counterterm. Since except for CKM matrix elements \cite{c10} all of the one-loop-level counterterms of physical parameters are real numbers and don't contain Heaviside function \cite{cin0}, the first kind self energy doesn't contribute to the singularities of the real part of the self energy, because except the one-loop-level counterterm the left part of this kind self energy is an one-loop-level self energy which real part doesn't contain singularities. Here we don't need to worry about the problem that the CKM matrix elements and their counterterms are complex numbers, because the total contribution of them to the real part of the gauge boson's self energy is real number (the correctness of this conclusion can be see from the following calculations). So we only need to calculate the contributions of the second kind self energy.

According to the {\em cutting rules} \cite{c8} the second kind self energy can be classified into three kinds: one kind doesn't contain singularity, the second kind contains singularities, but its singularities don't contribute to the real part of the self energy, the third kind contains singularities and its singularities contribute to the real part of the self energy. The topologies of the three kind self energies are shown in Fig.1, Fig.2 and Fig.3.
\begin{figure}[htbp]
\begin{center}
  \epsfig{file=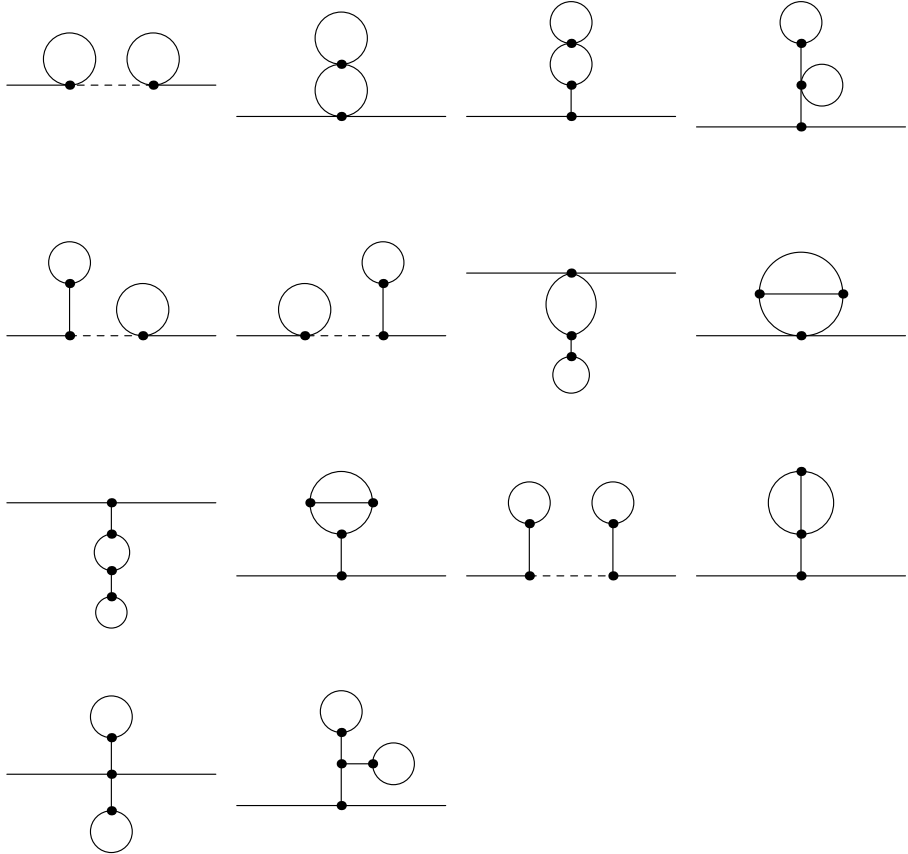, width=12cm} \\
  \caption{Topologies of the two-loop-level self energy which lacks counterterm and singularity.}
\end{center}
\end{figure}
\begin{figure}[htbp]
\begin{center}
  \epsfig{file=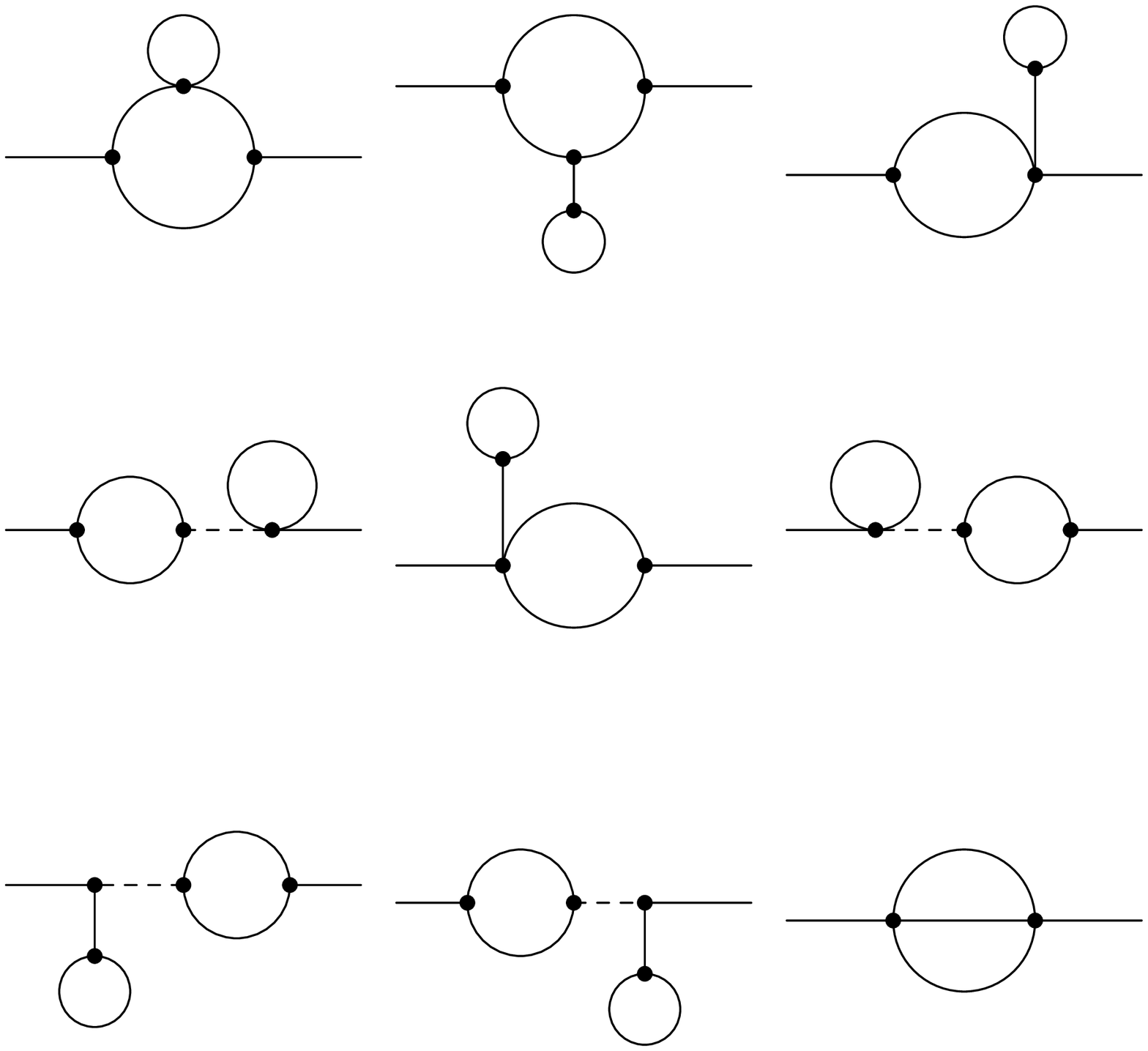, width=9.5cm} \\
  \caption{Topologies of the two-loop-level self energy which lacks counterterm and its singularities
  don't contribute to the real part of the self energy.}
\end{center}
\end{figure}
\begin{figure}[htbp]
\begin{center}
  \epsfig{file=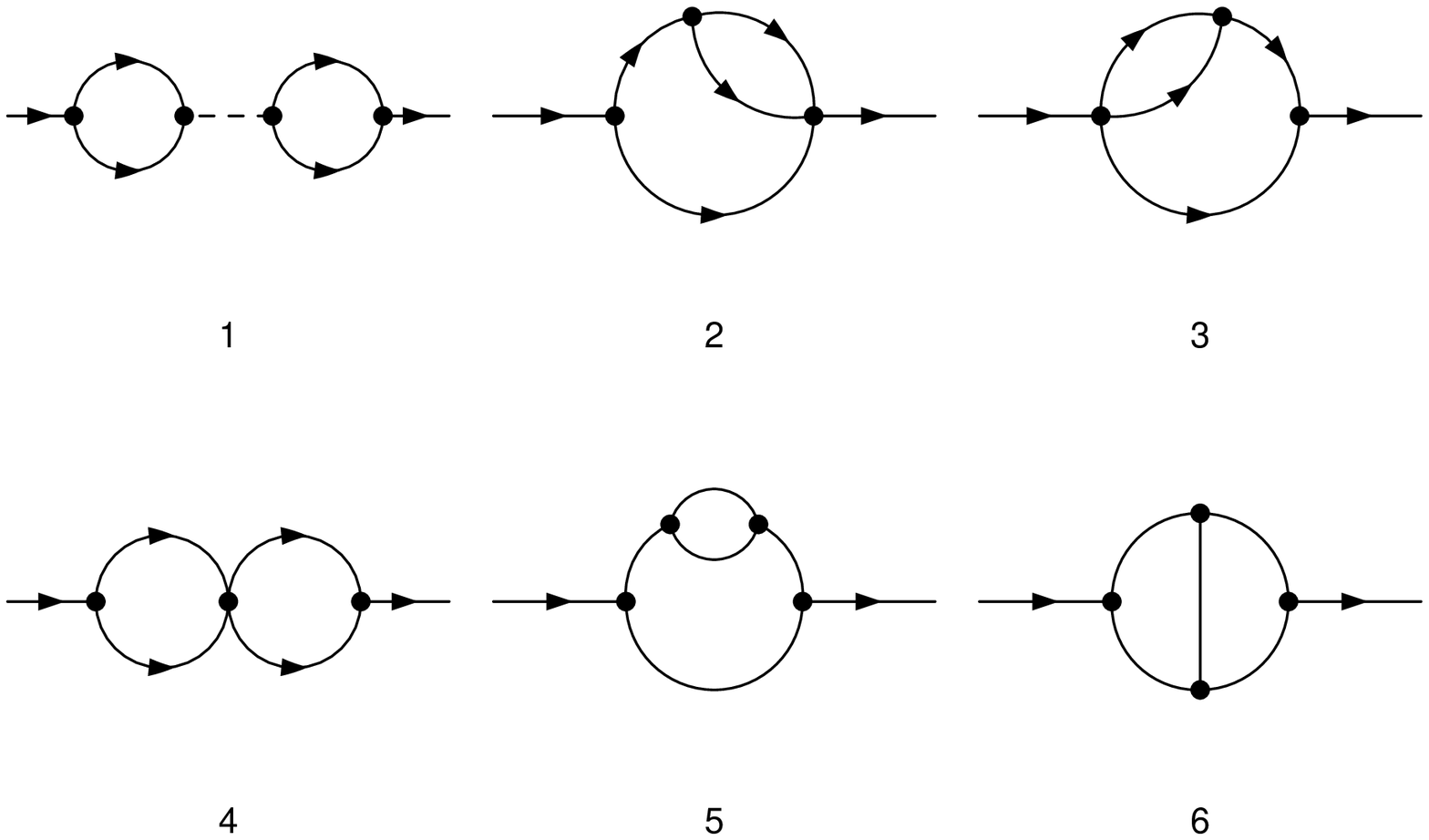, width=9.5cm} \\
  \caption{Topologies of the two-loop-level self energy which lacks counterterm and its singularities
  contribute to the real part of the self energy.}
\end{center}
\end{figure}
Here we note the middle propagator (denoted by broken line) in the one-particle-reducible diagrams of Fig.(1-3) is different from the external-line particles. The tadpole diagrams are also included in Fig.(1-3), because we don't introduce the tadpole counterterm \cite{c5}.

Obviously we only need to calculate the contribution of the singularities of Fig.3 to the real part of the gauge boson's self energy. In Fig.3 we also draw the possible cuts/singularities of the first four topologies which contribute to the real part of the gauge boson's self energy (the arrow on the inner line denotes the corresponding propagator is cut \cite{c8}). The possible cuts of the left two topologies which contribute to the real part of the gauge boson's self energy are shown in Fig.4 and Fig.5.
\begin{figure}[htbp]
\begin{center}
  \epsfig{file=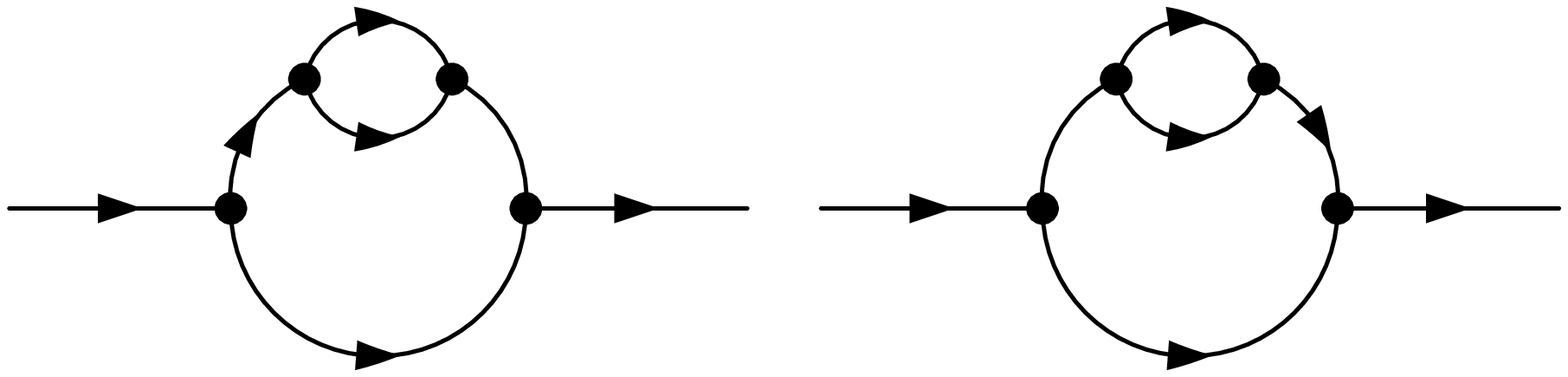, width=6.5cm} \\
  \caption{Possible cuts of the 5th topology of Fig.3 which contribute to the real part of the self energy.}
\end{center}
\end{figure}
\begin{figure}[htbp]
\begin{center}
  \epsfig{file=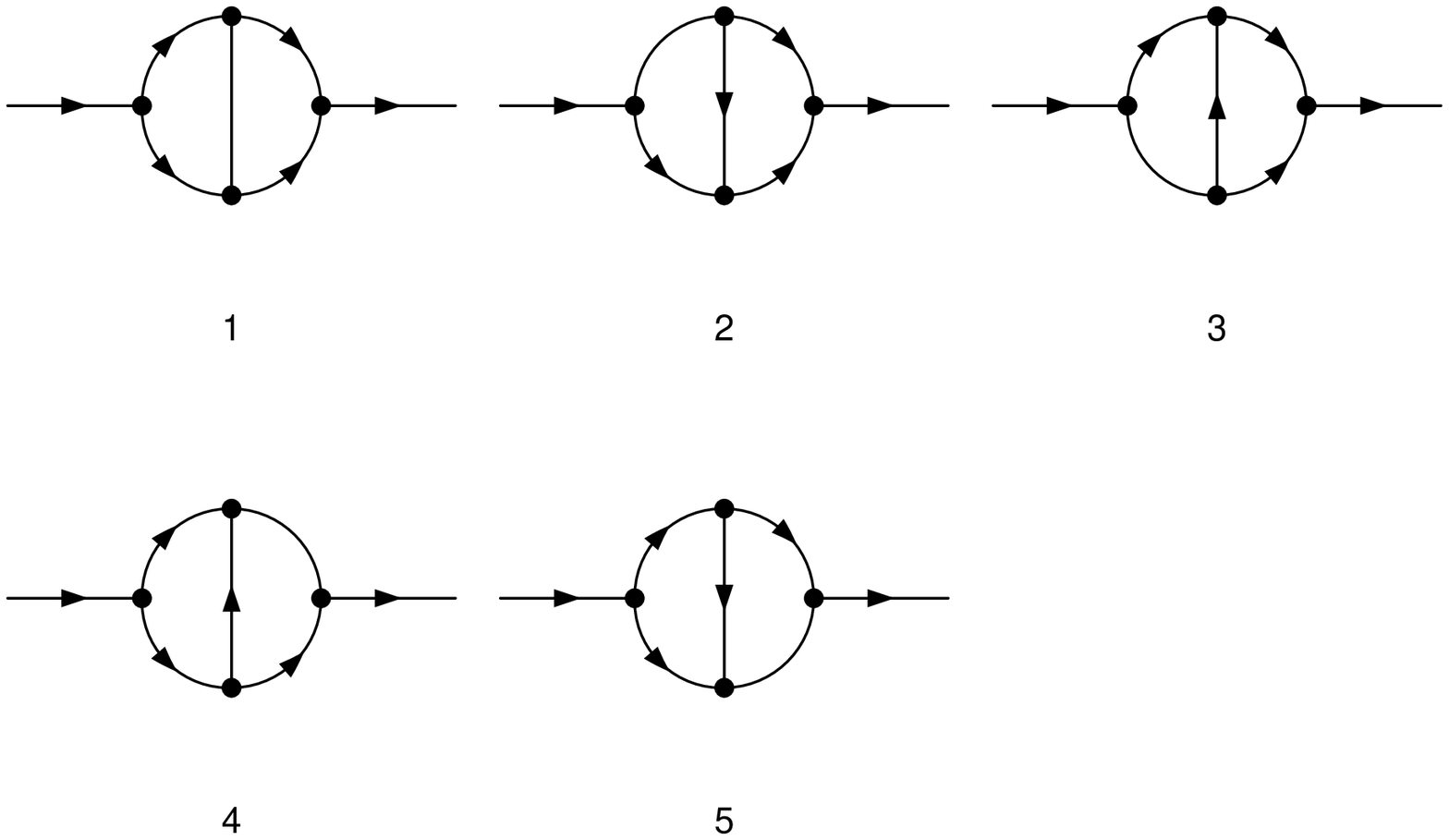, width=9.5cm} \\
  \caption{Possible cuts of the 6th topology of Fig.3 which contribute to the real part of the self energy.}
\end{center}
\end{figure}

\subsection{Gauge dependence of W mass counterterm under the two mass renormalization prescriptions}

In the standard model of particle physics the first topology of Fig.3 doesn't contribute to W transverse self energy, so we don't need to calculate its contribution. For the second topology of Fig.3 there are 39 Feynman diagrams in the standard model, but none of them satisfies the corresponding cutting condition. The case of the third topology of Fig.3 is same as the case of the second topology. For the 4th topology of Fig.3 there are two W self energy diagrams as shown in Fig.6 which satisfy the corresponding cutting condition.
\begin{figure}[htbp]
\begin{center}
  \epsfig{file=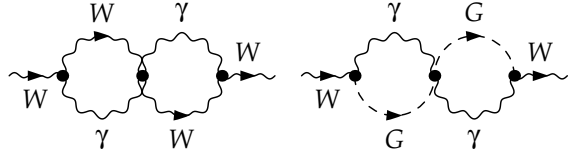, width=7.5cm} \\
  \caption{W self energy diagrams which satisfy the 4th topology of Fig.3 and the corresponding cutting condition.}
\end{center}
\end{figure}
Using the cutting rules we obtain the gauge-parameter-dependent contribution of the cuts of Fig.6 to the real part of W transverse self energy:
\beq
  Re\Sigma^T_{WW}(m_W^2)|_{\xi_W-cut}\,=\,\frac{\alpha^2 m_W^2}{4608}(1-\xi_W)^3
  (\xi_W^5-3\xi_W^4-6\xi_W^3-46\xi_W^2+165\xi_W+465)\,\theta[1-\xi_W]\,,
\eeq
where $\Sigma^T_{WW}$ is W transverse self energy, $m_W$ and $\xi_W$ is W's mass and gauge parameter, $\alpha$ is the fine structure constant, $\theta$ is the Heaviside function, and the subscript $\xi_W\hspace{-2mm}-\hspace{-1mm}cut$ denotes the $\xi_W$-dependent contribution from the cuts/singularities. In the follows we restrict ourselves to $\xi_W>0$ \cite{c3}.

For the 5th topology of Fig.3 there are 14 W's self energy diagrams as shown in Fig.7 which are $\xi_W$-dependent and satisfy the cutting conditions of Fig.4.
\begin{figure}[htbp]
\begin{center}
  \epsfig{file=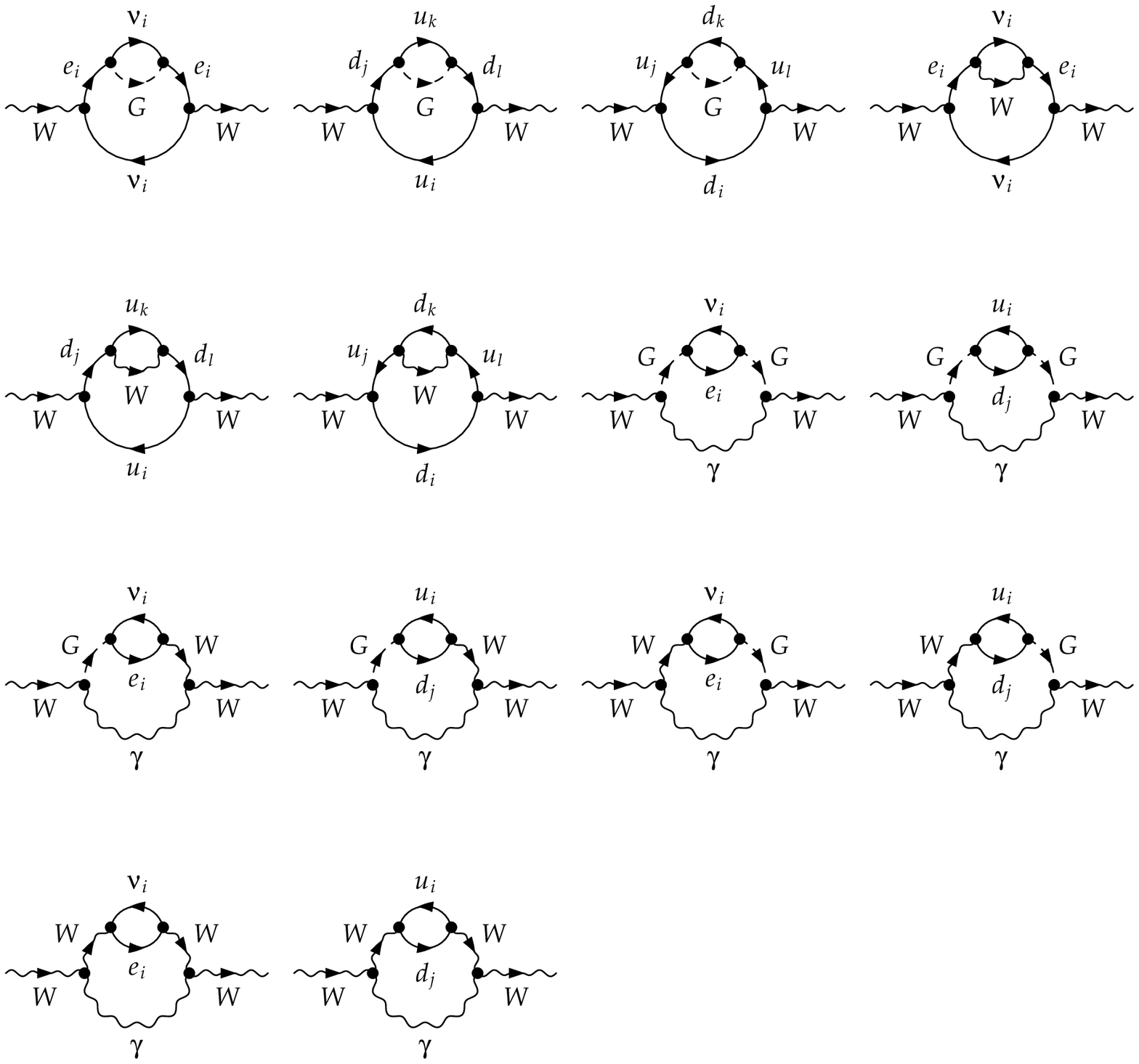, width=13cm} \\
  \caption{$\xi_W$-dependent W self energy diagrams which satisfy the 5th topology of
  Fig.3 and the cutting conditions of Fig.4.}
\end{center}
\end{figure}
After careful calculations we obtain the gauge-parameter-dependent contribution of the cuts of Fig.7 to the real part of W transverse self energy:
\beqa
  Re\Sigma^T_{WW}(m_W^2)|_{\xi_W-cut}&=&\frac{\alpha^2 m_W^2}{128 s_w^4}\Bigl{[}
  \sum_{i=e,\mu,\tau}\frac{1}{x_i}(1-x_i)(x_i-\xi_W)^2 (x_i^2+x_i-2)\,
  \theta[m_i-\sqrt{\xi_W}m_W] \nonumber \\
  &+&\frac{1}{\xi_W^2}s_w^2(1-\xi_W)^3\sum_{i=e,\mu,\tau}x_i(x_i-\xi_W)^2\,
  \theta[\sqrt{\xi_W}m_W-m_i]\,\theta[1-\xi_W] \nonumber \\
  &+&3\sum_{i=u,c}\sum_{j=d,s,b}\frac{1}{x_i}|V_{ij}|^2(\xi_W-x_i+x_j)
  A_{ij}\,B_{ij}\,C_{ij}\,\theta[m_i-m_j-\sqrt{\xi_W}m_W] \nonumber \\
  &+&3\sum_{i=u,c}\sum_{j=d,s,b}\frac{1}{x_j}|V_{ij}|^2(\xi_W-x_j+x_i)
  A_{ij}\,B_{ij}\,C_{ij}\,\theta[m_j-m_i-\sqrt{\xi_W}m_W] \nonumber \\
  &+&\frac{3}{\xi_W^2}s_w^2(1-\xi_W)^3\sum_{i=u,c}\sum_{j=d,s,b}|V_{ij}|^2
  (\xi_W(x_i+x_j)-(x_i-x_j)^2)\,C_{ij} \nonumber \\
  &\times&\theta[\sqrt{\xi_W}m_W-m_i-m_j]\,\theta[1-\xi_W]\,\Bigr{]} \,,
\eeqa
where $s_w$ is the sine of the weak mixing angle, $x_i=m_i^2/m_W^2$,  $x_j=m_j^2/m_W^2$, $V_{ij}$ is the CKM matrix element \cite{c10}, and
\beqa
  A_{ij}&=&\sqrt{(x_i-x_j)^2-2(x_i+x_j)+1}\,, \nonumber \\
  B_{ij}&=&2-(x_i+x_j)-(x_i-x_j)^2\,, \nonumber \\
  C_{ij}&=&\sqrt{(x_i-x_j)^2-2\xi_W(x_i+x_j)+\xi_W^2}\,.
\eeqa

For the 6th topology of Fig.3 there are 53 W self energy diagrams as shown in Fig.8 which are $\xi_W$-dependent and satisfy the cutting conditions of Fig.5.
\begin{figure}[htbp]
\begin{center}
  \epsfig{file=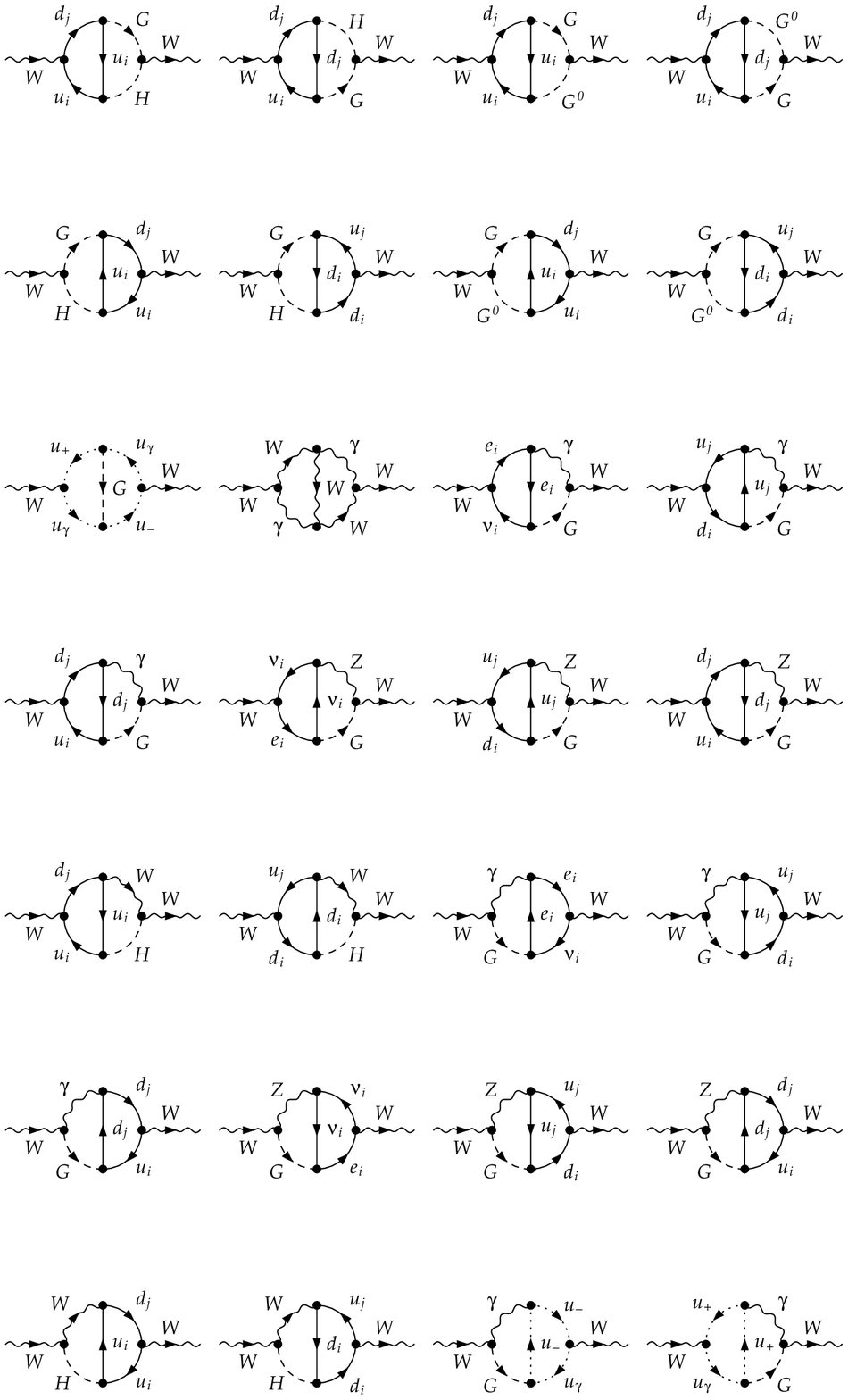, width=13cm}
\end{center}
\end{figure}
\begin{figure}[htbp]
\begin{center}
  \epsfig{file=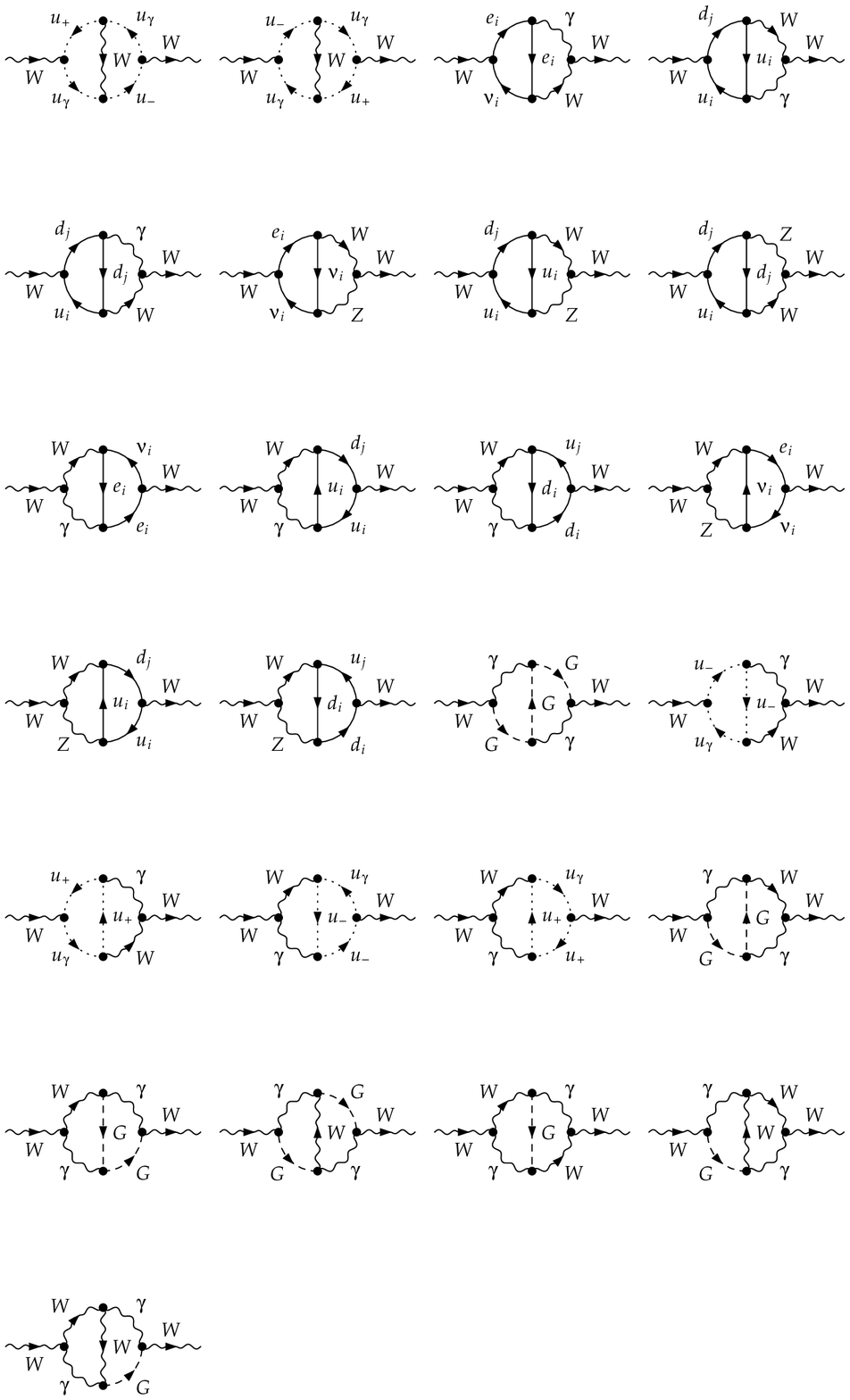, width=13cm} \\
  \caption{$\xi_W$-dependent W self energy diagrams which satisfy the 6th topology of
  Fig.3 and the cutting conditions of Fig.5.}
\end{center}
\end{figure}
We will calculate the contributions of the five cuts of Fig.5 one by one. Firstly we obtain the gauge-parameter-dependent contribution of the first cut of Fig.5 to the real part of W transverse self energy:
\beqa
  Re\Sigma^T_{WW}(m_W^2)|_{\xi_W-cut}&=&-\frac{\alpha^2 m_W^2}{576 s_w^2}\Bigl{[}
  3\sum_{i=u,c}\sum_{j=d,s,b}|V_{ij}|^2 A_{ij}\,B_{ij}+\sum_{i=e,\mu,\tau}(1-x_i)^2
  (2+x_i) \Bigr{]} \nonumber \\
  &\times&(1-\xi_W)(\xi_W^2-2\xi_W-11)\theta[1-\xi_W]-\frac{\alpha^2 m_W^2}{4608}(1-\xi_W)^3
  \nonumber \\
  &\times&(\xi_W^5-3\xi_W^4-6\xi_W^3-46\xi_W^2+165\xi_W+465)\,\theta[1-\xi_W]\,.
\eeqa
Then we obtain the gauge-parameter-dependent contributions of the second and third cuts of Fig.5 to the real part of W transverse self energy:
\beqa
  Re\Sigma^T_{WW}(m_W^2)|_{\xi_W-cut}&=&-\frac{\alpha^2 m_W^2}{256 s_w^4}\Bigl{[}
  \sum_{i=e,\mu,\tau}\frac{1}{x_i}(1-x_i)(x_i-\xi_W)^2 (x_i^2+x_i-2)\,
  \theta[m_i-\sqrt{\xi_W}m_W] \nonumber \\
  &+&\frac{1}{\xi_W^2}s_w^2(1-\xi_W)^3\sum_{i=e,\mu,\tau}x_i(x_i-\xi_W)^2\,
  \theta[\sqrt{\xi_W}m_W-m_i]\,\theta[1-\xi_W] \nonumber \\
  &+&3\sum_{i=u,c}\sum_{j=d,s,b}\frac{1}{x_i}|V_{ij}|^2(\xi_W-x_i+x_j)
  A_{ij}\,B_{ij}\,C_{ij}\,\theta[m_i-m_j-\sqrt{\xi_W}m_W] \nonumber \\
  &+&3\sum_{i=u,c}\sum_{j=d,s,b}\frac{1}{x_j}|V_{ij}|^2(\xi_W-x_j+x_i)
  A_{ij}\,B_{ij}\,C_{ij}\,\theta[m_j-m_i-\sqrt{\xi_W}m_W] \nonumber \\
  &+&\frac{3}{\xi_W^2}s_w^2(1-\xi_W)^3\sum_{i=u,c}\sum_{j=d,s,b}|V_{ij}|^2
  (\xi_W(x_i+x_j)-(x_i-x_j)^2)\,C_{ij} \nonumber \\
  &\times&\theta[\sqrt{\xi_W}m_W-m_i-m_j]\,\theta[1-\xi_W]\,\Bigr{]}\,.
\eeqa
Lastly we find the gauge-parameter-dependent contributions of the 4th and 5th cuts of Fig.5 to the real part of W transverse self energy are same as those of the second and the third cuts of Fig.5 (this point can be seen from the symmetries of the four cuts).

Summing up all of the above results we obtain the gauge dependence of the singularities of the real part of W two-loop-level transverse self energy (see Eqs.(6,7,9,10) and the corresponding discussions)
\beqa
  Re\Sigma^T_{WW}(m_W^2)|_{\xi_W-cut}&=&-\frac{\alpha^2 m_W^2}{576 s_w^2}
  \Bigl{[}3\sum_{i=u,c}\sum_{j=d,s,b}|V_{ij}|^2 A_{ij}\,B_{ij}
  +\sum_{i=e,\mu,\tau}(1-x_i)^2(2+x_i) \Bigr{]} \nonumber \\
  &\times&(1-\xi_W)(\xi_W^2-2\xi_W-11)\,\theta[1-\xi_W]\,.
\eeqa
From Eq.(5) one finds Eq.(11) is just the gauge dependence of the part containing Heaviside functions of W mass counterterm under the on-shell mass renormalization prescription. So Eq.(11) proves  the W mass counterterm of on-shell mass renormalization prescription is gauge dependent.

In order to discuss the gauge dependence of W mass counterterm of the pole mass renormalization prescription we calculate the term (see Eq.(5))
\beqa
  m_W\Gamma_W\,Im\,\Sigma_{WW}^{T\prime}(m_W^2)|_{\xi_W-cut}&=&\frac{\alpha^2 m_W^2}
  {576 s_w^2}\Bigl{[} 3\sum_{i=u,c}\sum_{j=d,s,b}|V_{ij}|^2 A_{ij}\,B_{ij}
  +\sum_{i=e,\mu,\tau}(1-x_i)^2(2+x_i) \Bigr{]} \nonumber \\
  &\times&(1-\xi_W)(\xi_W^2-2\xi_W-11)\,\theta[1-\xi_W]\,.
\eeqa
Combining Eq.(11) and Eq.(12) one gets (see Eq.(5))
\beq
  \delta m_W^2|_{\xi_W-cut}\,=\,0 \hspace{12mm}
  under\hspace{2mm}pole\hspace{2mm}mass\hspace{2mm}renormalization\hspace{2mm}prescription \,.
\eeq
This result indicates the part containing Heaviside functions of W mass counterterm of the pole mass renormalization prescription is gauge independent.

\subsection{Gauge dependence of Z mass counterterm under the two mass renormalization prescriptions}

Similarly as the case of W gauge boson We only calculate the gauge dependence of the part containing Heaviside function of the real part of Z two-loop-level transverse self energy. The topologies of Z two-loop-level self energy needing calculated have been shown in Fig.3.

For the first topology of Fig.3 only the diagram whose middle
propagator (denoted by the broken line) is photon contributes to Z
transverse self energy. After careful calculation we obtain the
$\xi_W$-dependent contribution of the cut of the first topology of
Fig.3 to the real part of Z transverse self energy \beqa
  &&Re\Sigma^T_{ZZ}(m_Z^2)|_{\xi_W-cut} \nonumber \\
  =&&\frac{\alpha^2 m_W^2}{6912\,c_w^6\,s_w^2}
  (1-4\,c_w^2\,\xi_W)^{3/2}\Bigl{[}\frac{3}{c_w^2}(1-4\,c_w^2\,\xi_W)^{3/2}
  +8\bigl{(} 3(4 c_w^2-3)\sum_{i=e,\mu,\tau}+2(8 c_w^2-5)\sum_{i=u,c} \nonumber \\
  +&&(4 c_w^2-1)\sum_{i=d,s,b} \bigl{)}\sqrt{1-4\,c_w^2\,x_i}\,(2\,c_w^2\,x_i+1) \Bigr{]}\,
  \theta[\frac{1}{c_w}-2\sqrt{\xi_W}]+\frac{\alpha^2 m_W^2}{1728\,c_w^6}D\,E \nonumber \\
  \times&&\Bigl{[} \frac{3\,s_w^2}{c_w^2}D\,E
  -\frac{3}{c_w^2}(1-4\,c_w^2\,\xi_W)^{3/2}-4\bigl{(} 3(4 c_w^2-3)\sum_{i=e,\mu,\tau}
  +2(8 c_w^2-5)\sum_{i=u,c}+(4 c_w^2-1)\sum_{i=d,s,b} \bigl{)} \nonumber \\
  \times&&\sqrt{1-4\,c_w^2\,x_i}\,(2\,c_w^2\,x_i+1) \Bigr{]}\,
  \theta[\frac{1}{c_w}-\sqrt{\xi_W}-1]\,,
\eeqa
where $m_Z$ is Z mass, $c_w$ is the cosine of the weak mixing angle, and
\beqa
  D&=&\sqrt{(\xi_W-1)^2 c_w^4-2(\xi_W+1)c_w^2+1}\,, \nonumber \\
  E&=&(\xi_W-1)^2 c_w^4-2(\xi_W-5)c_w^2+1\,.
\eeqa

For the second topology of Fig.3 there are four Z self-energy diagrams as shown in Fig.9 which satisfy the corresponding cutting condition.
\begin{figure}[htbp]
\begin{center}
  \epsfig{file=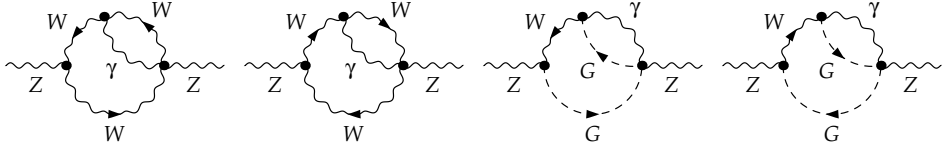, width=12.5cm} \\
  \caption{Z self-energy diagrams which satisfy the second topology of
  Fig.3 and the corresponding cutting condition.}
\end{center}
\end{figure}
By the cutting rules we obtain the $\xi_W$-dependent contribution
of the cuts of Fig.9 to the real part of Z transverse self energy
: \beqa
  Re\Sigma^T_{ZZ}(m_Z^2)|_{\xi_W-cut}&=&\frac{\alpha^2 m_W^2}{1536\,c_w^4}
  (\xi_W-1)D\bigl{[} (\xi_W-1)^3(\xi_W^3-\xi_W^2-3\xi_W-33)c_w^6 \nonumber \\
  &-&(\xi_W-1)(3\xi_W^4-9\xi_W^3-29\xi_W^2+101\xi_W+366)c_w^4 \nonumber \\
  &+&(3\xi_W^4-10\xi_W^3-22\xi_W^2+170\xi_W-93)c_w^2-\xi_W^3 \nonumber \\
  &+&2\xi_W^2+5\xi_W-18 \bigr{]}\theta[\frac{1}{c_w}-\sqrt{\xi_W}-1]\,.
\eeqa
For the third topology of Fig.3 there are also four Z self-energy diagrams as shown in Fig.10 which satisfy the corresponding cutting condition.
\begin{figure}[htbp]
\begin{center}
  \epsfig{file=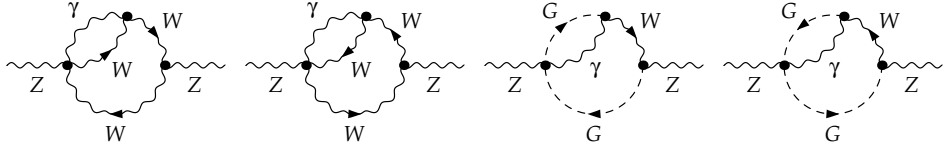, width=12.5cm} \\
  \caption{Z self-energy diagrams which satisfy the third topology of
  Fig.3 and the corresponding cutting condition.}
\end{center}
\end{figure}
Obviously Fig.9 and Fig.10 are right-and-left symmetric. Through calculations we find the $\xi_W$-dependent contribution of the cuts of Fig.10 to the real part of Z transverse self energy is just equal to that of Fig.9.

For the 4th topology of Fig.3 there are six Z self-energy diagrams as shown in Fig.11 which satisfy the corresponding cutting rules.
\begin{figure}[htbp]
\begin{center}
  \epsfig{file=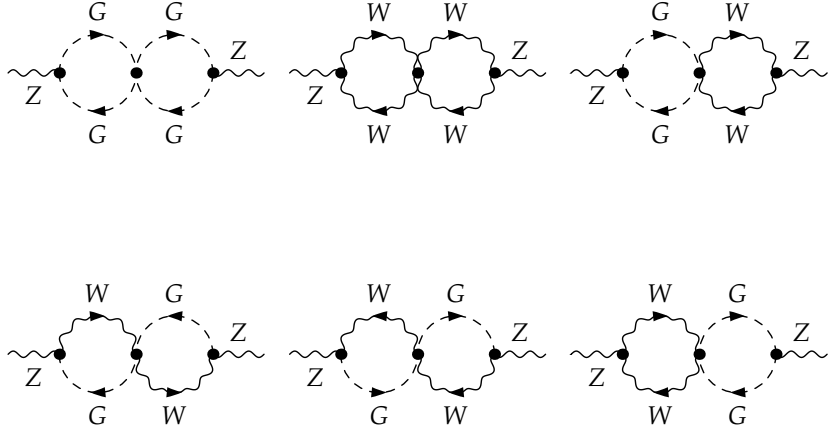, width=11cm} \\
  \caption{Z self-energy diagrams which satisfy the 4th topology of
  Fig.3 and the corresponding cutting condition.}
\end{center}
\end{figure}
After careful calculations we obtain the $\xi_W$-dependent contribution of the cuts of Fig.11 to the real part of Z transverse self energy:
\beqa
  Re\Sigma^T_{ZZ}(m_Z^2)|_{\xi_W-cut}&=&\frac{\alpha^2 m_W^2}{4608\,c_w^8\,s_w^4}
  (1-4\,c_w^2\,\xi_W)^3
  (2 c_w^6-4 c_w^4+2 c_w^2-3)\,\theta[\frac{1}{c_w}-2\sqrt{\xi_W}] \nonumber \\
  &+&\frac{\alpha^2 m_W^2}{2304\,c_w^8}\bigl{[} (\xi_W-1)^6 c_w^{14}-6(\xi_W-1)^4
  (\xi_W^2+11\xi_W+22)c_w^{12} \nonumber \\
  &+&3(\xi_W-1)^2(12\xi_W^3+65\xi_W^2+10\xi_W+201)c_w^{10} \nonumber \\
  &-&2(45\xi_W^4+46\xi_W^3-228\xi_W^2-150\xi_W+415)c_w^8 \nonumber \\
  &+&3(40\xi_W^3-19\xi_W^2-98\xi_W+109)c_w^6-6(15\xi_W^2-17\xi_W-12)c_w^4 \nonumber \\
  &+&(36\xi_W-35)c_w^2-6 \bigr{]}\,\theta[\frac{1}{c_w}-\sqrt{\xi_W}-1]
  -\frac{\alpha^2 m_W^2}{1152\,c_w^8\,s_w^2}D\,\sqrt{1-4\,c_w^2\,\xi_W} \nonumber \\
  &\times&\bigl{[}4(\xi_W-1)^2\xi_W\,c_w^{10}-(4\xi_w^3+\xi_W^2-38\xi_W+1)c_w^8 \nonumber \\
  &+&3(4\xi_W^3+19\xi_W^2-32\xi_W-3)c_w^6-3(9\xi_W^2-10\xi_w-8)c_w^4 \nonumber \\
  &+&(18\xi_w-11)c_w^2-3 \bigr{]}\,\theta[\frac{1}{c_w}-\sqrt{\xi_W}-1]\,.
\eeqa

For the 5th topology of Fig.3 there are 84 Z self-energy diagrams as shown in Fig.12 which satisfy the cutting conditions of Fig.4.
\begin{figure}[htbp]
\begin{center}
  \epsfig{file=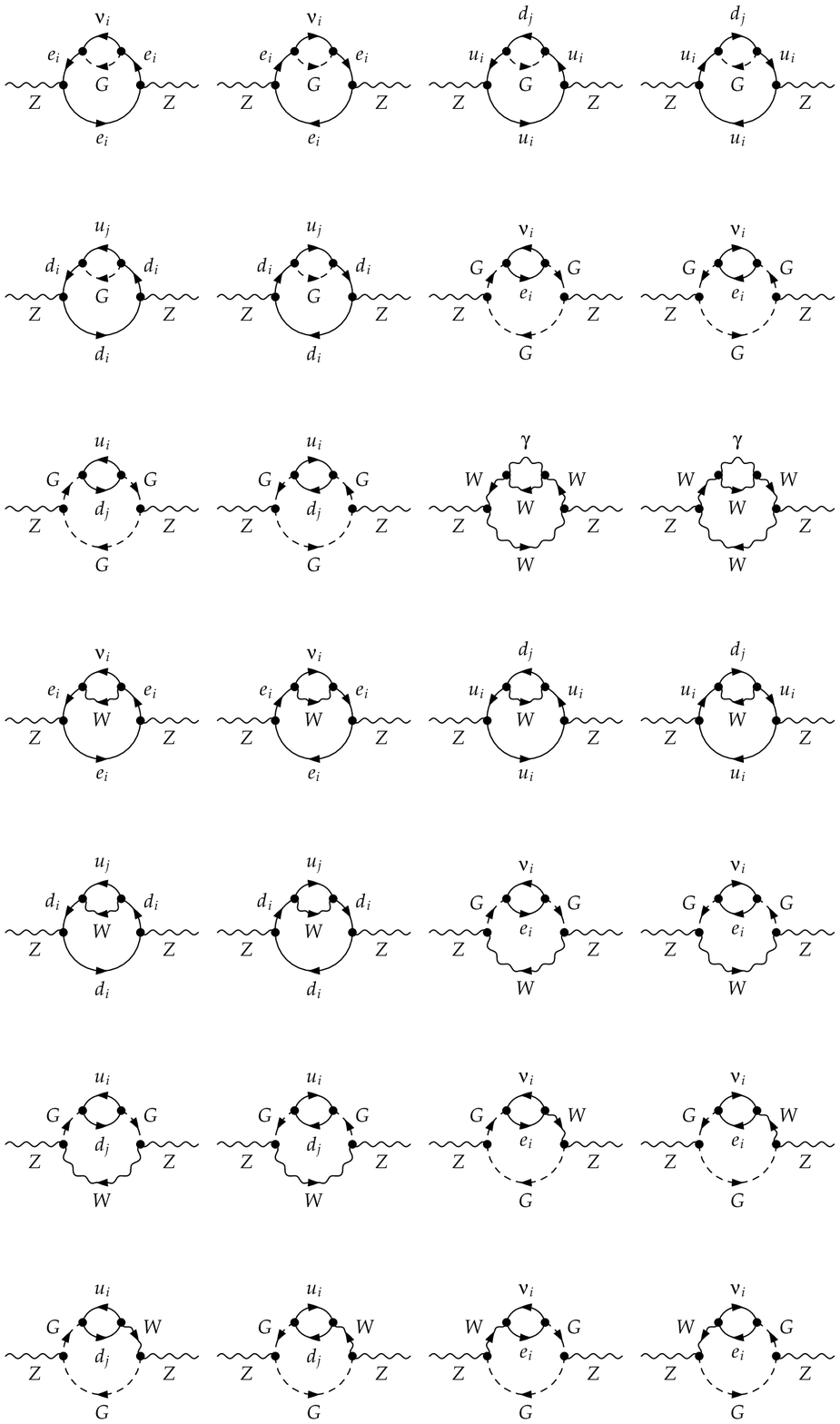, width=13cm}
\end{center}
\end{figure}
\begin{figure}[htbp]
\begin{center}
  \epsfig{file=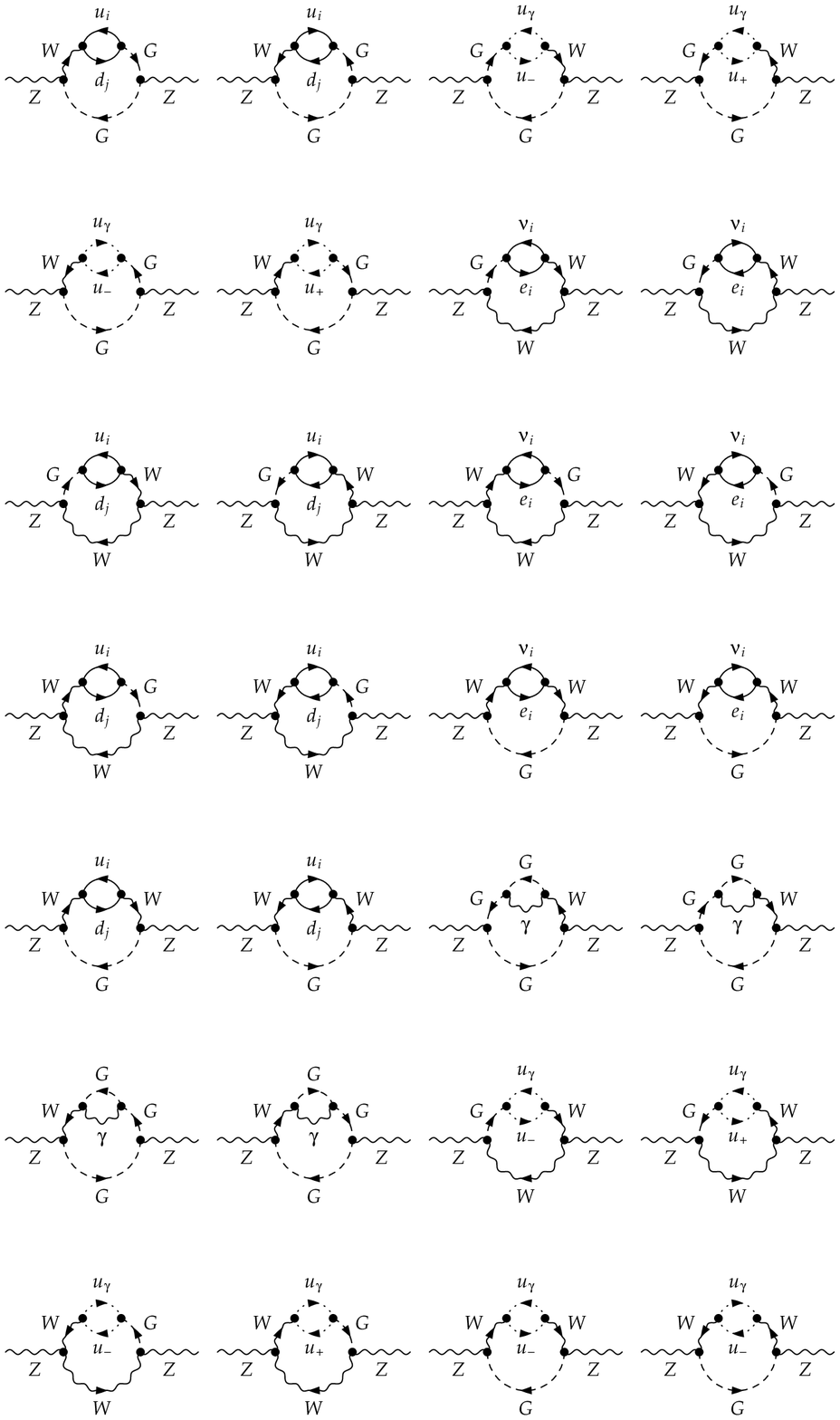, width=13cm}
\end{center}
\end{figure}
\begin{figure}[htbp]
\begin{center}
  \epsfig{file=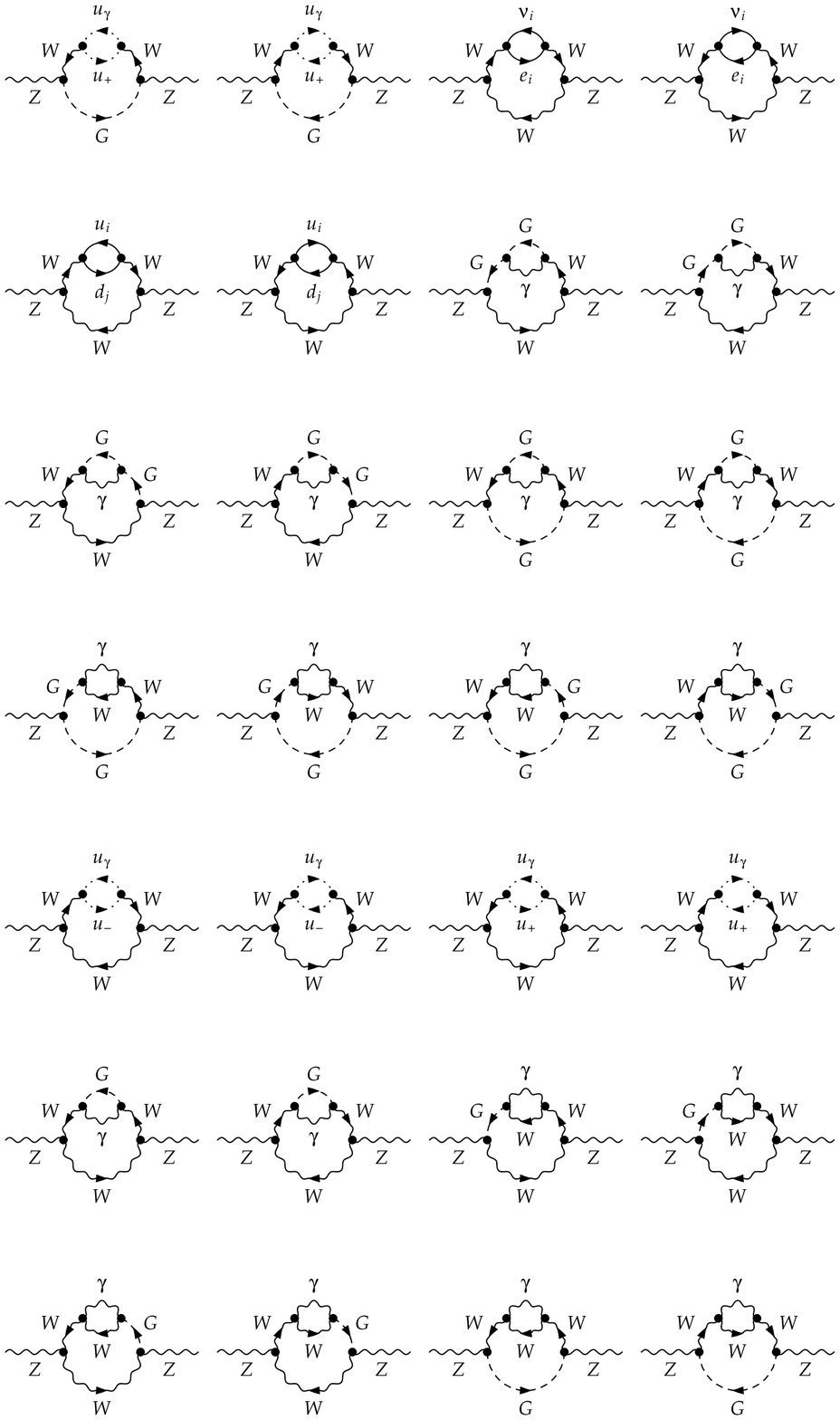, width=13cm} \\
  \caption{Z self-energy diagrams which satisfy the 5th topology of
  Fig.3 and the cutting conditions of Fig.4.}
\end{center}
\end{figure}
After careful calculations we obtain the $\xi_W$-dependent contribution of the cuts of Fig.12 to the real part of Z transverse self energy:
\beqa
  &&Re\Sigma^T_{ZZ}(m_Z^2)|_{\xi_W-cut} \nonumber \\
  =&&\frac{\alpha^2 m_W^2}{192\,c_w^2\,s_w^2}D\,E\bigl{[} \sum_{i=e,\mu,\tau}
  (1-x_i)^2(2+x_i)+3\sum_{i=u,c}\sum_{j=d,s,b}|V_{ij}|^2 A_{ij}\,B_{ij} \bigr{]}\,
  \theta[\frac{1}{c_w}-\sqrt{\xi_W}-1] \nonumber \\
  +&&\frac{\alpha^2 m_W^2}{64\,s_w^4\,\xi_W^2}(1-4\,c_w^2\,\xi_W)^{3/2}
  \sum_{i=e,\mu,\tau}x_i(x_i-\xi_W)^2\,\theta[\frac{1}{c_w}-2\sqrt{\xi_W}]\,
  \theta[\sqrt{\xi_W}m_W-m_i] \nonumber \\
  +&&\frac{3\,\alpha^2 m_W^2}{64\,s_w^4\,\xi_W^2}(1-4\,c_w^2\,\xi_W)^{3/2}
  \sum_{i=u,c}\sum_{j=d,s,b}|V_{ij}|^2\,C_{ij}\,(\xi_W(x_i+x_j)-(x_i-x_j)^2) \nonumber \\
  \times&&\theta[\frac{1}{c_w}-2\sqrt{\xi_W}]\,\theta[\sqrt{\xi_W}m_W-m_i-m_j]
  -\frac{\alpha^2 m_W^2}{64\,c_w^4\,s_w^4}(2 c_W^2-1)\sum_{i=e,\mu,\tau}
  \frac{1}{x_i}\sqrt{1-4\,c_w^2\,x_i} \nonumber \\
  \times&&(x_i-\xi_W)^2(2 c_w^2-1+c_w^2(4 c_w^2-5)x_i)\,
  \theta[m_i-\sqrt{\xi_W}m_W] \nonumber \\
  +&&\frac{\alpha^2 m_W^2}{192\,c_w^4\,s_w^4}(4 c_W^2-1)\sum_{i=u,c}\sum_{j=d,s,b}
  \frac{1}{x_i}|V_{ij}|^2\,C_{ij}\,\sqrt{1-4\,c_w^2\,x_i}\,(\xi_W-x_i+x_j) \nonumber \\
  \times&&(4 c_w^2-1+c_w^2(8 c_w^2-11)x_i)\,
  \theta[m_i-m_j-\sqrt{\xi_W}m_W] \nonumber \\
  +&&\frac{\alpha^2 m_W^2}{192\,c_w^4\,s_w^4}
  (2 c_W^2+1)\sum_{i=u,c}\sum_{j=d,s,b}
  \frac{1}{x_j}|V_{ij}|^2\,C_{ij}\,\sqrt{1-4\,c_w^2\,x_j}
  (\xi_W-x_j+x_i) \nonumber \\
  \times&&(2 c_w^2+1+c_w^2(4 c_w^2-7)x_j)\,
  \theta[m_j-m_i-\sqrt{\xi_W}m_W]\,.
\eeqa

For the 6th topology of Fig.3 there are 124 Z self-energy diagrams as shown in Fig.13 which are $\xi_W$ dependent and satisfy the cutting conditions of Fig.5.
\begin{figure}[htbp]
\begin{center}
  \epsfig{file=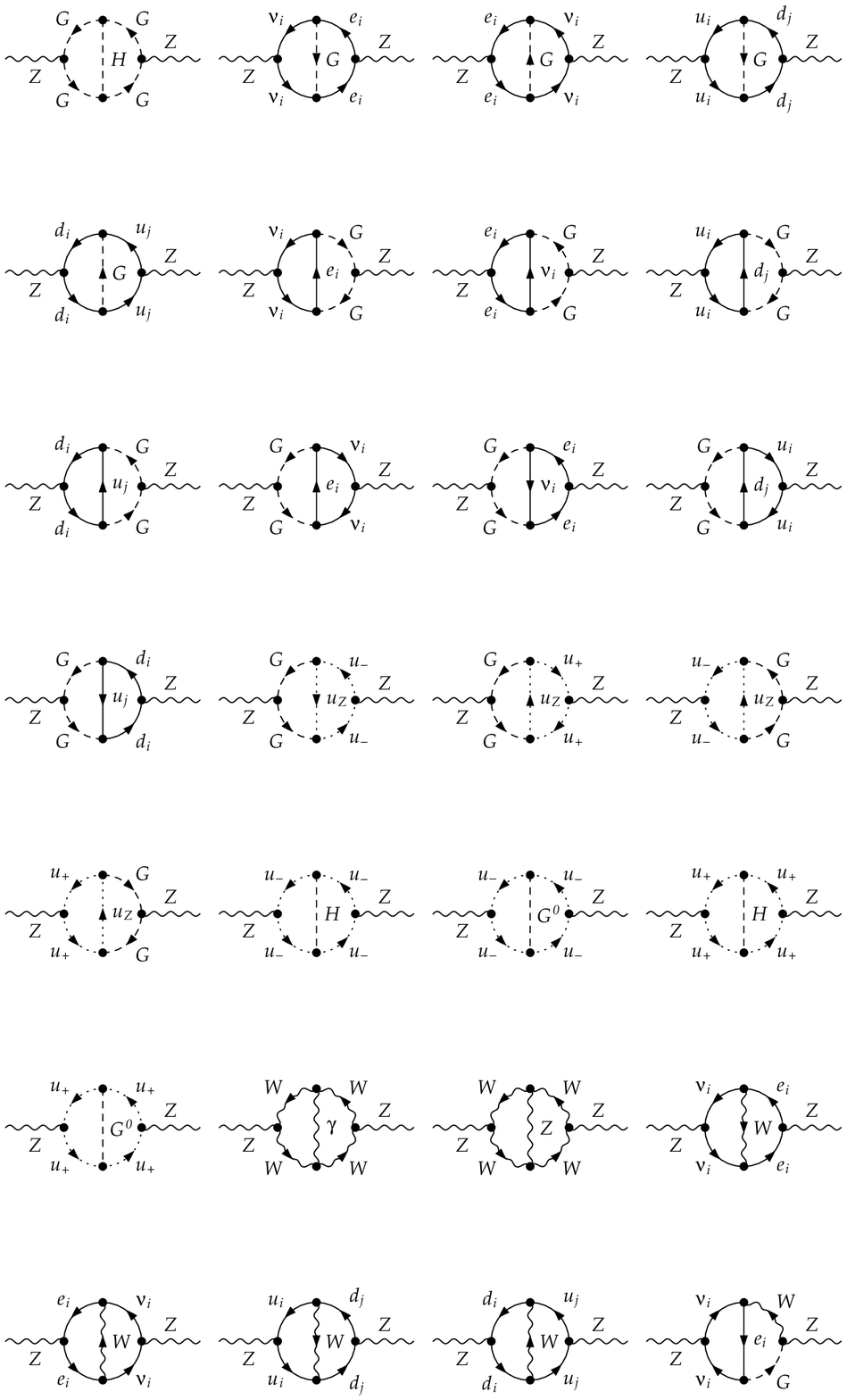, width=13cm}
\end{center}
\end{figure}
\begin{figure}[htbp]
\begin{center}
  \epsfig{file=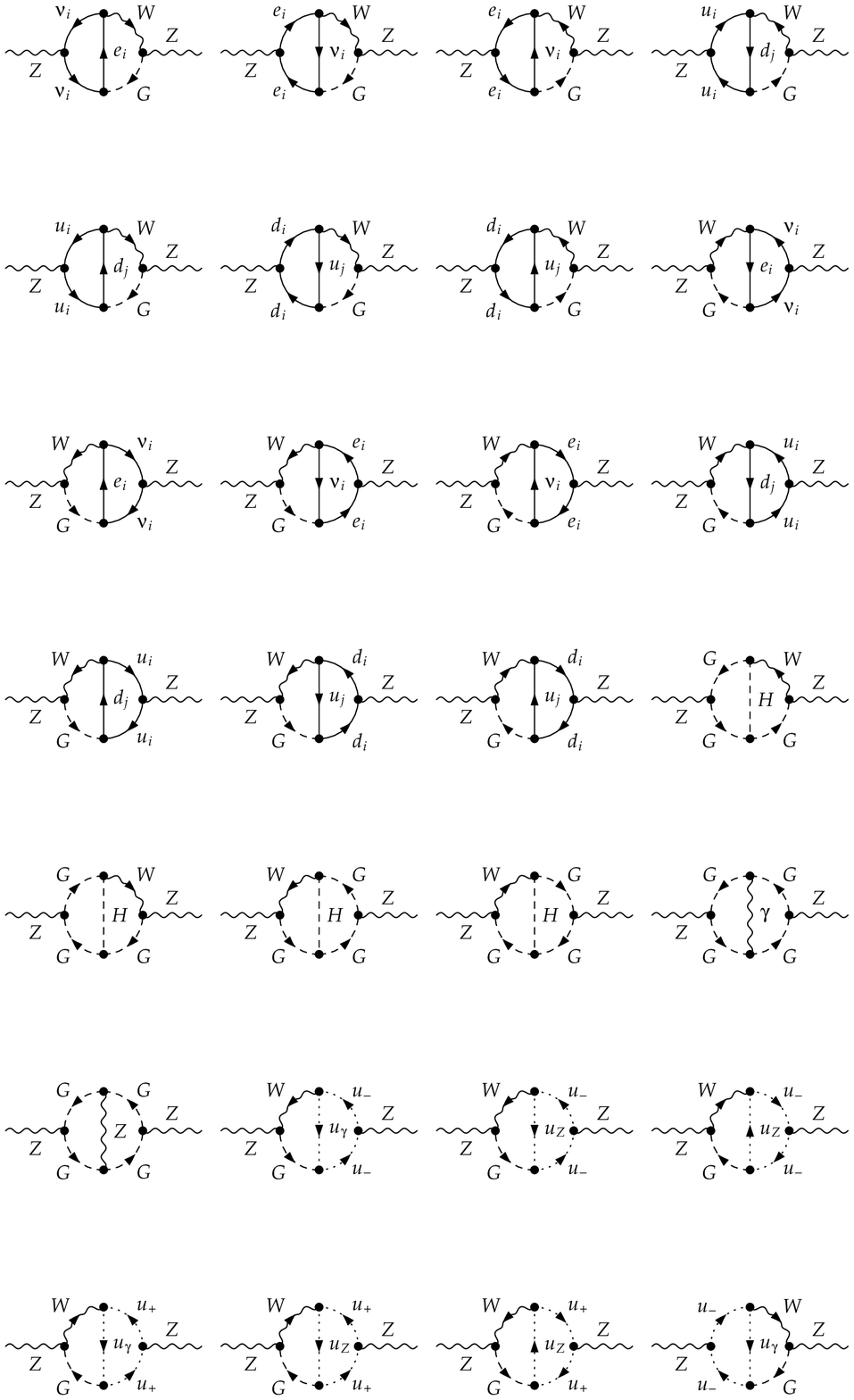, width=13cm}
\end{center}
\end{figure}
\begin{figure}[htbp]
\begin{center}
  \epsfig{file=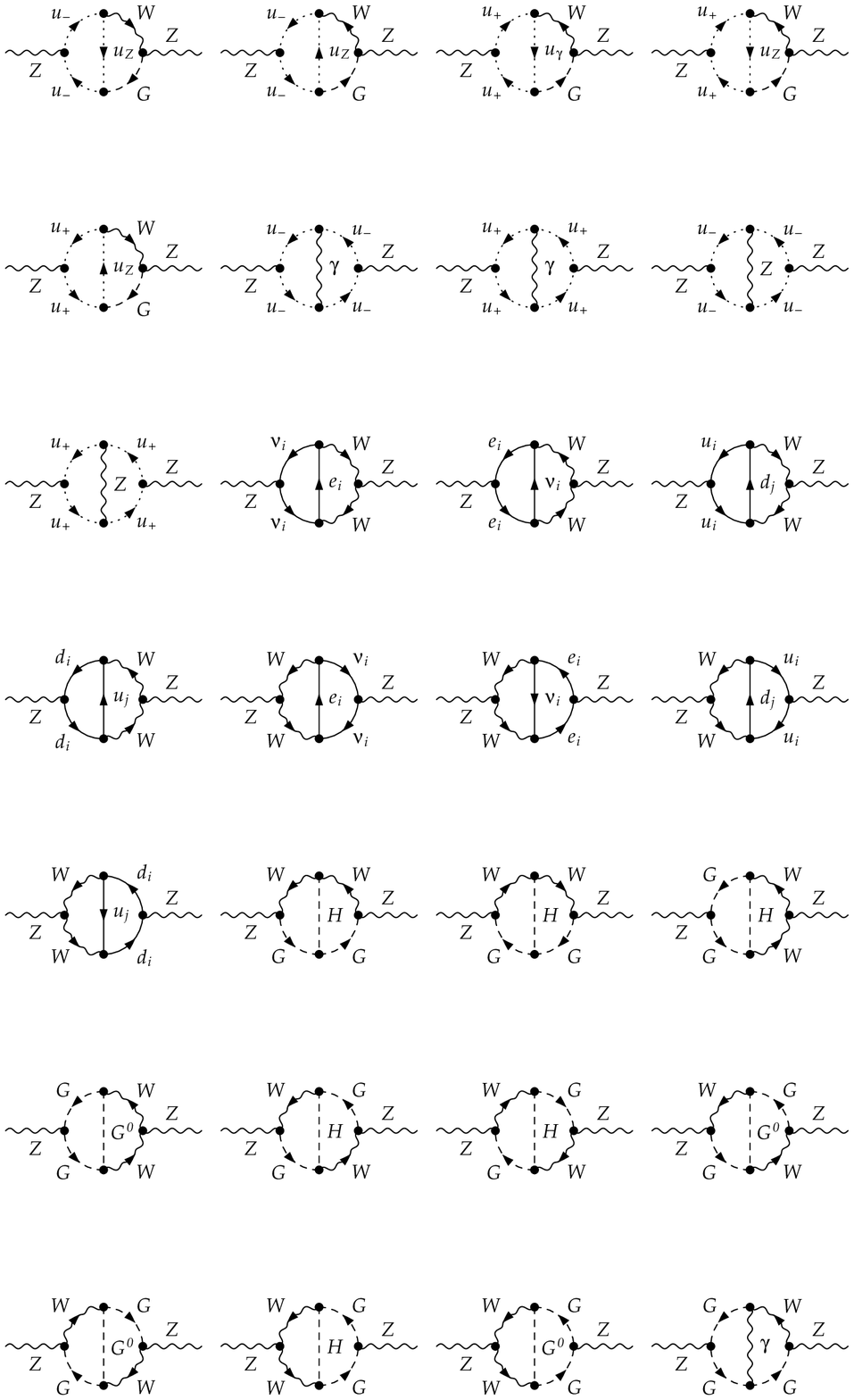, width=13cm}
\end{center}
\end{figure}
\begin{figure}[htbp]
\begin{center}
  \epsfig{file=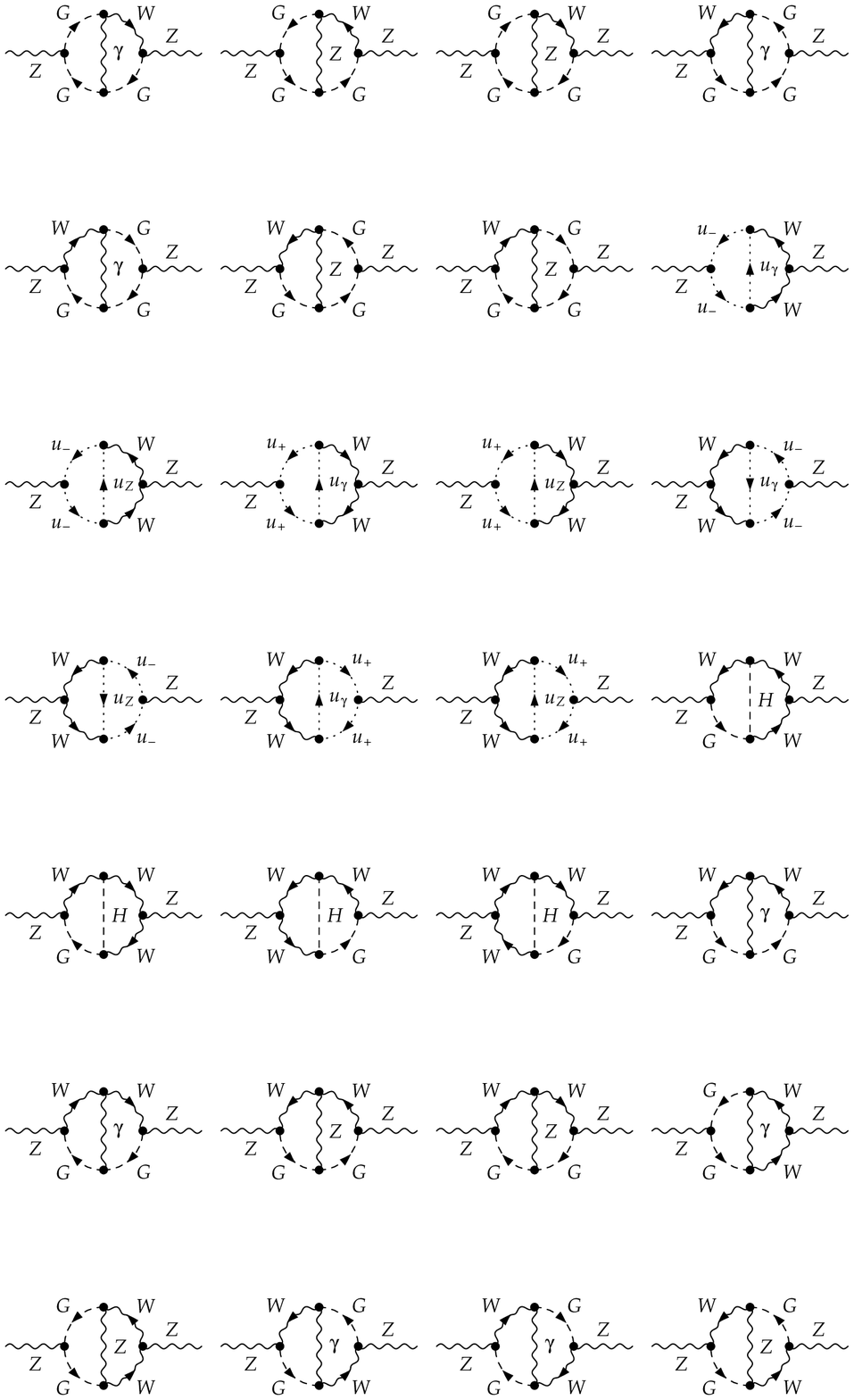, width=13cm}
\end{center}
\end{figure}
\begin{figure}[htbp]
\begin{center}
  \epsfig{file=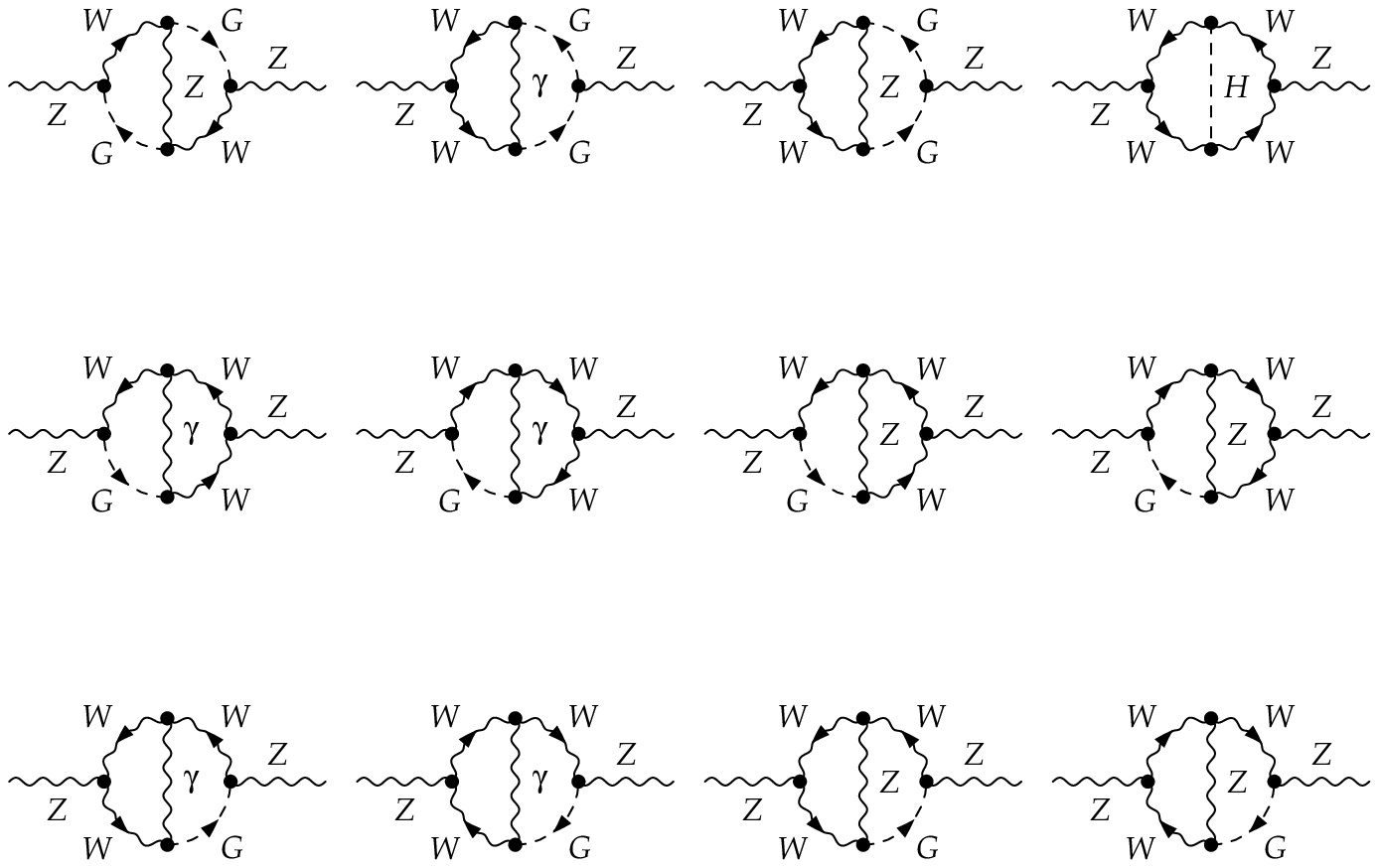, width=13cm} \\
  \caption{$\xi_W$-dependent Z self-energy diagrams which satisfy the 6th topology of
  Fig.3 and the cutting conditions of Fig.5.}
\end{center}
\end{figure}
We will calculate the contributions of the five cuts of Fig.5 one by one. Firstly we obtain the $\xi_W$-dependent contribution of the first cut of Fig.5 to the real part of Z transverse self energy:
\beqa
  Re\Sigma^T_{ZZ}(m_Z^2)|_{\xi_W-cut}&=&\frac{\alpha^2 m_W^2}
  {4608\,c_w^8\,s_w^4}(1-4\,c_w^2\,\xi_W)^3(1+4 c_w^4-2 c_w^6)\,
  \theta[\frac{1}{c_w}-2\sqrt{\xi_W}] \nonumber \\
  &-&\frac{\alpha^2 m_W^2}{576\,c_w^6\,s_w^4}(1-4\,c_w^2\,\xi_W)^{3/2}\bigl{[}
  3+\bigl{(} \sum_{i=e,\mu,\tau}2 c_w^2-1+c_w^2(4 c_w^2-5)x_i \nonumber \\
  &+&\sum_{i=u,c}4 c_w^2-1+c_w^2(8 c_w^2-11)x_i
  +\sum_{i=d,s,b}2 c_w^2+1+c_w^2(4 c_W^2-7)x_i \bigr{)} \nonumber \\
  &\times&\sqrt{1-4\,c_w^2\,x_i}\,\bigr{]}\,\theta[\frac{1}{c_w}-2\sqrt{\xi_W}]
  +\frac{\alpha^2 m_W^2}{1152\,c_w^8\,s_w^2}D\,\sqrt{1-4\,c_w^2\,\xi_W} \nonumber \\
  &\times&\bigl{[} 4(\xi_W-1)^2\xi_W c_w^{10}+(4\xi_W^3-17\xi_W^2+46\xi_W-1)c_w^8 \nonumber \\
  &+&(4\xi_W^3+55\xi_W^2-20\xi_W-11)c_w^6-(9\xi_W^2+42\xi_W-6)c_w^4 \nonumber \\
  &+&(6\xi_W+7)c_W^2-1 \bigr{]}\,\theta[\frac{1}{c_W}-\sqrt{\xi_W}-1] \nonumber \\
  &+&\frac{\alpha^2 m_W^2}{2304\,c_w^8}\bigl{[} 3(\xi_W-1)^6 c_w^{14}+
  2(\xi_W-1)^4(\xi_W^2+25\xi_W+100)c_w^{12} \nonumber \\
  &-&3(\xi_W-1)^2(4\xi_W^3+85\xi_W^2+58\xi_W+141)c_w^{10} \nonumber \\
  &+&6(5\xi_W^4+74\xi_W^3-144\xi_W^2-50\xi_W-13)c_w^8 \nonumber \\
  &-&(40\xi_W^3+411\xi_W^2-798\xi_W-581)c_w^6+6(5\xi_W^2+31\xi_W-42)c_w^4 \nonumber \\
  &-&3(4\xi_W+11)c_w^2+2 \bigr{]}\,\theta[\frac{1}{c_w}-\sqrt{\xi_W}-1]
  +\frac{\alpha^2 m_W^2}{288\,c_w^6\,s_w^2}D\,E \nonumber \\
  &\times&\bigl{[} 3+\bigl{(} \sum_{i=e,\mu,\tau}(2 c_w^2-1+c_w^2(4 c_w^2-5)x_i)
  +\sum_{i=u,c}(4 c_w^2-1+c_w^2(8 c_w^2-11)x_i) \nonumber \\
  &+&\sum_{i=d,s,b}(2 c_w^2+1+c_w^2(4 c_W^2-7)x_i) \bigr{)}
  \sqrt{1-4\,c_w^2\,x_i}\,\bigr{]}\,\theta[\frac{1}{c_w}-\sqrt{\xi_W}-1]\,.
\eeqa
Then we obtain the $\xi_W$-dependent contribution of the second and third cuts of Fig.5 to the real part of Z transverse self energy:
\beqa
  &&Re\Sigma^T_{ZZ}(m_Z^2)|_{\xi_W-cut} \nonumber \\
  =&&-\frac{\alpha^2 m_W^2}{384\,c_w^2\,s_w^2}D\,E\bigl{[} \sum_{i=e,\mu,\tau}
  (1-x_i)^2(2+x_i)+3\sum_{i=u,c}\sum_{j=d,s,b}|V_{ij}|^2 A_{ij}\,B_{ij} \bigr{]}\,
  \theta[\frac{1}{c_w}-\sqrt{\xi_W}-1] \nonumber \\
  -&&\frac{\alpha^2 m_W^2}{1536\,c_w^4}(\xi_W-1)D\bigl{[} (\xi_W-1)^3
  (\xi_W^3-\xi_W^2-3\xi_W-33)c_w^6 \nonumber \\
  -&&(\xi_W-1)(3\xi_W^4-9\xi_W^3-29\xi_W^2+101\xi_W+366)c_w^4 \nonumber \\
  +&&(3\xi_W^4-10\xi_W^3-22\xi_W^2+170\xi_W-93)c_w^2
  -\xi_W^3+2\xi_W^2+5\xi_W-18 \bigr{]}\theta[\frac{1}{c_w}-\sqrt{\xi_W}-1] \nonumber \\
  -&&\frac{\alpha^2 m_W^2}{128\,s_w^4\,\xi_W^2}(1-4\,c_w^2\,\xi_W)^{3/2}
  \sum_{i=e,\mu,\tau}x_i(x_i-\xi_W)^2\,\theta[\frac{1}{c_w}-2\sqrt{\xi_W}]\,
  \theta[\sqrt{\xi_W}m_W-m_i] \nonumber \\
  -&&\frac{3\,\alpha^2 m_W^2}{128\,s_w^4\,\xi_W^2}(1-4\,c_w^2\,\xi_W)^{3/2}
  \sum_{i=u,c}\sum_{j=d,s,b}|V_{ij}|^2\,C_{ij}\,(\xi_W(x_i+x_j)-(x_i-x_j)^2)
  \nonumber \\
  \times&&\theta[\frac{1}{c_w}-2\sqrt{\xi_W}]\,\theta[\sqrt{\xi_W}m_W-m_i-m_j]
  +\frac{\alpha^2 m_W^2}{128\,c_w^4\,s_w^4}(2 c_W^2-1) \nonumber \\
  \times&&\sum_{i=e,\mu,\tau}
  \frac{1}{x_i}\sqrt{1-4\,c_w^2\,x_i}\,(x_i-\xi_W)^2(2 c_w^2-1+c_w^2(4 c_w^2-5)x_i)\,
  \theta[m_i-\sqrt{\xi_W}m_W] \nonumber \\
  -&&\frac{\alpha^2 m_W^2}{384\,c_w^4\,s_w^4}(4 c_W^2-1)\sum_{i=u,c}\sum_{j=d,s,b}
  \frac{1}{x_i}|V_{ij}|^2\,C_{ij}\,\sqrt{1-4\,c_w^2\,x_i}\,(\xi_W-x_i+x_j) \nonumber \\
  \times&&(4 c_w^2-1+c_w^2(8 c_w^2-11)x_i)\,\theta[m_i-m_j-\sqrt{\xi_W}m_W]
  \nonumber \\
  -&&\frac{\alpha^2 m_W^2}{384\,c_w^4\,s_w^4}
  (2 c_W^2+1)\sum_{i=u,c}\sum_{j=d,s,b}\frac{1}{x_j}|V_{ij}|^2\,C_{ij}\,
  \sqrt{1-4\,c_w^2\,x_j}(\xi_W-x_j+x_i) \nonumber \\
  \times&&(2 c_w^2+1+c_w^2(4 c_w^2-7)x_j)\,
  \theta[m_j-m_i-\sqrt{\xi_W}m_W]\,.
\eeqa
From Fig.5 one readily sees the 4th and 5th cuts are right-and-left symmetric with the second and third cuts. After careful calculations we also find the $\xi_W$-dependent contribution of the 4th and 5th cuts of Fig.5 to the real part of Z transverse self energy is equal to that of the second and third cuts of Fig.5.

Summing up all of the above results we obtain the gauge dependence of the part containing Heaviside functions of the real part of Z two-loop-level transverse self energy (see Eqs.(14-20) and the corresponding discussions)
\beqa
  &&Re\Sigma^T_{ZZ}(m_Z^2)|_{\xi_W-cut} \nonumber \\
  =&&\frac{\alpha^2 m_W^2}{1728\,c_w^6\,s_w^2}
  \Bigl{[} 9+3\sum_{i=e,\mu,\tau}\sqrt{1-4\,c_w^2\,x_i}\,
  (16 x_i\,c_w^6+(8-24 x_i)c_w^4+(7 x_i-12)c_w^2+5) \nonumber \\
  +&&\sum_{i=u,c}\sqrt{1-4\,c_w^2\,x_i}\,(64 x_i\,c_w^6+(32-80 x_i)c_w^4+
  (7 x_i-40)c_w^2+17) \nonumber \\
  +&&\sum_{i=d,s,b}\sqrt{1-4\,c_w^2\,x_i}\,(16 x_i\,c_w^6+8(1-x_i)c_w^4-(17 x_i+4)c_w^2+5)
  \Bigr{]} \nonumber \\
  \times&&\bigl{[} 2 D\,E\,\theta[\frac{1}{c_w}-\sqrt{\xi_W}-1]-\frac{1}{s_w^2}
  (1-4\,c_w^2\,\xi_W)^{3/2}\theta[\frac{1}{c_w}-2\sqrt{\xi_W}]\,\bigr{]}\,.
\eeqa
This result proves that the part containing Heaviside functions of Z mass counterterm is gauge dependent under the on-shell mass renormalization prescription, i.e. the Z mass counterterm is gauge dependent under the on-shell mass renormalization prescription.

In order to calculate the gauge dependence of Z mass definition of the pole mass renormalization prescription we need to calculate the following term (see Eq.(5)):
\beqa
  &&m_Z\Gamma_Z\,Im\,\Sigma_{ZZ}^{T\prime}(m_Z^2)|_{\xi_W-cut} \nonumber \\
  =&&\frac{\alpha^2 m_W^2}{1728\,c_w^6\,s_w^2}\Bigl{[} 9+3\sum_{i=e,\mu,\tau}
  \sqrt{1-4\,c_w^2\,x_i}\,(16 x_i\,c_w^6+(8-24 x_i)c_w^4+(7 x_i-12)c_w^2+5) \nonumber \\
  +&&\sum_{i=u,c}\sqrt{1-4\,c_w^2\,x_i}\,(64 x_i\,c_w^6+(32-80 x_i)c_w^4+
  (7 x_i-40)c_w^2+17) \nonumber \\
  +&&\sum_{i=d,s,b}\sqrt{1-4\,c_w^2\,x_i}\,(16 x_i\,c_w^6+8(1-x_i)c_w^4-(17 x_i+4)c_w^2+5)
  \Bigr{]} \nonumber \\
  \times&&\bigl{[} \frac{1}{s_w^2}(1-4\,c_w^2\,\xi_W)^{3/2}
  \theta[\frac{1}{c_w}-2\sqrt{\xi_W}]-2 D\,E\,\theta[\frac{1}{c_w}-\sqrt{\xi_W}-1]\,\bigr{]}\,.
\eeqa
From Eq.(5) and Eqs.(21,22) we get the gauge dependence of the part containing Heaviside functions of Z mass counterterm under the pole mass renormalization prescription:
\beq
  \delta m_Z^2|_{\xi_W-cut}\,=\,0 \hspace{12mm}
  under\hspace{2mm}pole\hspace{2mm}mass\hspace{2mm}renormalization\hspace{2mm}prescription \,.
\eeq

\subsection{Gauge dependence of the counterterm of the sine of the weak mixing angle under the two mass renormalization prescriptions}

From the two-loop-level W and Z's mass counterterms we can calculate the two-loop-level counterterm of the sine of the weak mixing angle. To two-loop level one has \cite{cin0}
\beq
  \delta s_w\,=\,\frac{c_w^2}{2 s_w}(\frac{\delta m_Z^2}{m_Z^2}-
  \frac{\delta m_W^2}{m_W^2})+\frac{c_w^2}{2 s_w}(\frac{\delta m_Z^2\,\delta m_W^2}
  {m_Z^2\,m_W^2}-\frac{(\delta m_Z^2)^2}{m_Z^4})-\frac{c_w^4}{8 s_w^3}
  (\frac{\delta m_Z^2}{m_Z^2}-\frac{\delta m_W^2}{m_W^2})^2+O(g^6)\,.
\eeq
The one-loop-level W and Z's mass counterterms have been proven gauge independent \cite{c5}. So we only need to calculate the gauge dependence of the first term of the r.h.s. of Eq.(24). From Eqs.(11,21) we obtain the gauge dependence of the part containing Heaviside functions of the two-loop-level $\delta s_w$ under the on-shell mass renormalization prescription
\beqa
  \delta s_w|_{\xi_W-cut}&=&\frac{\alpha^2}{3456\,c_w^2\,s_w^3}
  \Bigl{[} 9+3\sum_{i=e,\mu,\tau}\sqrt{1-4\,c_w^2\,x_i}\,
  (16 x_i\,c_w^6+(8-24 x_i)c_w^4+(7 x_i-12)c_w^2+5) \nonumber \\
  &+&\sum_{i=u,c}\sqrt{1-4\,c_w^2\,x_i}\,(64 x_i\,c_w^6+(32-80 x_i)c_w^4+
  (7 x_i-40)c_w^2+17) \nonumber \\
  &+&\sum_{i=d,s,b}\sqrt{1-4\,c_w^2\,x_i}\,(16 x_i\,c_w^6+8(1-x_i)c_w^4-(17 x_i+4)c_w^2+5)
  \Bigr{]} \nonumber \\
  &\times&\bigl{[} 2 D\,E\,\theta[\frac{1}{c_w}-\sqrt{\xi_W}-1]-\frac{1}{s_w^2}
  (1-4\,c_w^2\,\xi_W)^{3/2}\theta[\frac{1}{c_w}-2\sqrt{\xi_W}]\,\bigr{]} \nonumber \\
  &+&\frac{\alpha^2\,c_w^2}{1152\,s_w^3}\Bigl{[}3\sum_{i=u,c}\sum_{j=d,s,b}|V_{ij}|^2 A_{ij}\,B_{ij}
  +\sum_{i=e,\mu,\tau}(1-x_i)^2(2+x_i) \Bigr{]} \nonumber \\
  &\times&(1-\xi_W)(\xi_W^2-2\xi_W-11)\,\theta[1-\xi_W] \hspace{12mm}under\hspace{2mm}
  on\hspace{-1mm}-\hspace{-1mm}shell\hspace{2mm}prescription\,.
\eeqa
Eq.(25) implies $\delta s_w$ is gauge dependent under the on-shell mass renormalization prescription. On the other hand, from Eqs.(13,23) we obtain the gauge dependence of the part containing Heaviside functions of the two-loop-level $\delta s_w$ under the pole mass renormalization prescription
\beq
  \delta s_w|_{\xi_W-cut}\,=\,0 \hspace{12mm}under\hspace{2mm}pole\hspace{2mm}prescription\,.
\eeq

\section{Gauge dependence of physical result under the on-shell and pole mass renormalization prescriptions}

From the results of section II we have found the counterterms of W
and Z's mass and the sine of the weak mixing angle are gauge
dependent under the on-shell mass renormalization prescription,
but those gauge dependencies don't appear in the counterterms of
the pole mass renormalization prescription. Maybe this conclusion
is not enough to judge which renormalization prescription is
reasonable. So we will judge the reasonableness of the two renormalization prescriptions from the gauge
independence of physical result.

For example we calculate the gauge dependence of the two-loop-level cross section of the physical process $\mu\rightarrow\nu_{\mu}e^{-}\bar{\nu}_{e}$ under the two mass renormalization prescriptions. Note that we only calculate the gauge dependence of the part containing the Heaviside functions $\theta[1-\xi_W]$, $\theta[1/c_w-\sqrt{\xi_W}-1]$ and $\theta[1/c_w-2\sqrt{\xi_W}]$ of the cross section of the physical process. This will not affect our conclusion. Under this consideration only the diagrams containing the two-loop-level counterterms $\delta s_w$ and $\delta m_W^2$ as shown in Fig.14 need to be calculated.
\begin{figure}[htbp]
\begin{center}
  \epsfig{file=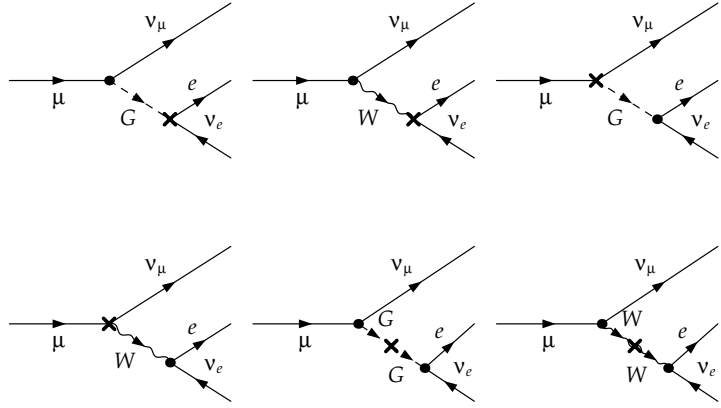, width=9.5cm} \\
  \caption{Diagrams of $\mu\rightarrow \nu_{\mu}e^{-}\bar{\nu}_{e}$ containing the two-loop-level
  counterterms $\delta s_w$ and $\delta m_W^2$.}
\end{center}
\end{figure}
This is because: 1) all of the one-loop-level physical parameter's counterterms and the two-loop-level counterterms of the lepton masses and electron charge don't contain the Heaviside functions $\theta[1-\xi_W]$, $\theta[1/c_w-\sqrt{\xi_W}-1]$ and $\theta[1/c_w-2\sqrt{\xi_W}]$; 2) the energy of the incoming particle of this process is order of muon energy which doesn't reach the threshold of the singularities containing the Heaviside functions $\theta[1-\xi_W]$, $\theta[1/c_w-\sqrt{\xi_W}-1]$ and $\theta[1/c_w-2\sqrt{\xi_W}]$, thus all of loop momentum integrals of the Feynman diagrams don't contribute to these Heaviside functions. We can easily get the contribution of Fig.14 to the physical amplitude $\mu\rightarrow \nu_{\mu}e^{-}\bar{\nu}_{e}$
\beqa
  {\cal M}(\mu\rightarrow\nu_{\mu} e^{-}\bar{\nu}_e)&\rightarrow&\frac{4\pi\,\alpha
  (m_e m_{\mu}F_1-m_W^2 F_2)}{m_W^2 s_w^3(m_W^2-m_e^2-2 q_2\cdot q_3)}
  \delta s_w+\frac{2\pi\alpha\,\delta m_W^2}{m_W^4 s_w^2(m_W^2-m_e^2-2 q_2\cdot q_3)^2}
  \nonumber \\
  &\times&\bigl{[} m_e m_{\mu}(2m_W^2-m_e^2-2 q_2\cdot q_3)F_1-m_W^4 F_2 \bigr{]}\,,
\eeqa
where $m_e$ and $m_{\mu}$ is the mass of electron and muon, $q_2$ and $q_3$ are the momentums of electron and the anti electron neutrino, and
\beq
  F_1\,=\,\bar{u}(q_1)\gamma_R u(p)\,\bar{u}(q_2)\gamma_L\nu(q_3)\,, \hspace{6mm}
  F_2\,=\,\bar{u}(q_1)\gamma^{\mu}\gamma_L u(p)\,
  \bar{u}(q_2)\gamma_{\mu}\gamma_L\nu(q_3)\,,
\eeq
where $p$ and $q_1$ are the momentums of muon and muon neutrino, and $\gamma_L$ and $\gamma_R$ are the left- and right- handed helicity operators. The contribution of Eq.(27) to the two-loop-level cross section of $\mu\rightarrow\nu_{\mu}e^{-}\bar{\nu}_e$ is
\beqa
  \sigma(\mu\rightarrow\nu_{\mu}e^{-}\bar{\nu}_e)&\propto&
  |{\cal M}(\mu\rightarrow\nu_{\mu}e^{-}\bar{\nu}_e)|^2 \nonumber \\
  &\rightarrow&\frac{16\pi^2\alpha^2 q_1\cdot q_2(m_e^2- m_{\mu}^2+2 q_1\cdot q_2)}{m_W^4 s_w^4}
  \bigl{(} \frac{2\delta s_w}{s_w}+\frac{\delta m_W^2}{m_W^2} \bigr{)}\,.
\eeqa
In Eq.(29) we have averaged the result over the incoming fermion's helicity states and summed up the results for the different outgoing fermions' helicity states. On the other hand we only keep the lowest order of the quantities $m_e^2/m_W^2$, $m_{\mu}^2/m_W^2$ and so on in Eq.(29), since the energies of the external-line particles are very small compared with $m_W^2$.

From Eqs.(11,25) and Eq.(29) we obtain the gauge dependence of the part containing the Heaviside functions $\theta[1-\xi_W]$, $\theta[1/c_w-\sqrt{\xi_W}-1]$ and $\theta[1/c_w-2\sqrt{\xi_W}]$ of the two-loop-level cross section of $\mu\rightarrow\nu_{\mu}e^{-}\bar{\nu}_e$ under the on-shell mass renormalization prescription
\beqa
  \sigma_2(\mu\rightarrow\nu_{\mu}e^{-}\bar{\nu}_e)_{\xi_W-cut}&\rightarrow&
  \frac{\pi^2\alpha^4 q_1\cdot q_2(m_e^2- m_{\mu}^2+2 q_1\cdot q_2)}
  {108 m_W^4\,c_w^2\,s_w^8}\Bigl{[} 9 \nonumber \\
  &+&3\sum_{i=e,\mu,\tau}\sqrt{1-4\,c_w^2\,x_i}\,
  (16 x_i\,c_w^6+(8-24 x_i)c_w^4+(7 x_i-12)c_w^2+5) \nonumber \\
  &+&\sum_{i=u,c}\sqrt{1-4\,c_w^2\,x_i}\,(64 x_i\,c_w^6+(32-80 x_i)c_w^4+
  (7 x_i-40)c_w^2+17) \nonumber \\
  &+&\sum_{i=d,s,b}\sqrt{1-4\,c_w^2\,x_i}\,(16 x_i\,c_w^6+8(1-x_i)c_w^4-(17 x_i+4)c_w^2+5)
  \Bigr{]} \nonumber \\
  &\times&\bigl{[} 2 D\,E\,\theta[\frac{1}{c_w}-\sqrt{\xi_W}-1]-\frac{1}{s_w^2}
  (1-4\,c_w^2\,\xi_W)^{3/2}\theta[\frac{1}{c_w}-2\sqrt{\xi_W}]\,\bigr{]} \nonumber \\
  &+&\frac{\pi^2\alpha^4 q_1\cdot q_2(m_e^2- m_{\mu}^2+2 q_1\cdot q_2)}{36 m_W^4\,s_w^8}
  (2 c_w^2-1)(1-\xi_W)(\xi_W^2-2\xi_W-11) \nonumber \\
  &\times&\Bigl{[}3\sum_{i=u,c}\sum_{j=d,s,b}|V_{ij}|^2 A_{ij}\,B_{ij}
  +\sum_{i=e,\mu,\tau}(1-x_i)^2(2+x_i) \Bigr{]}\,\theta[1-\xi_W]\,.
\eeqa
Eq.(30) implies the on-shell mass renormalization prescription makes the cross section of the physical process $\mu\rightarrow\nu_{\mu}e^{-}\bar{\nu}_e$ gauge dependent. So the on-shell mass renormalization prescription is a wrong mass renormalization prescription beyond one-loop level. The quantitative order of this gauge dependence can be seen in Fig.15.
\begin{figure}[htbp]
\begin{center}
  \epsfig{file=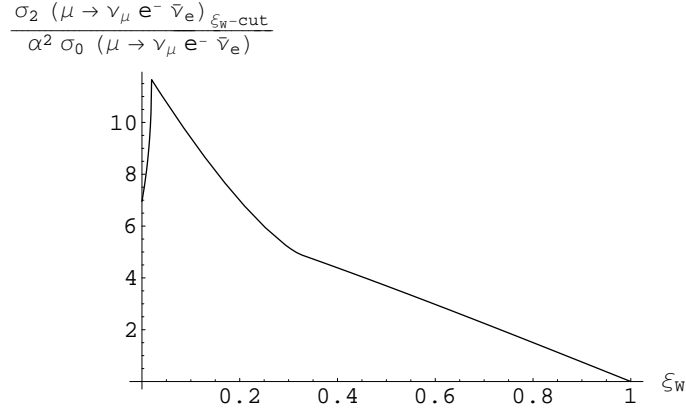, width=9cm} \\
  \caption{Gauge dependence of the two-loop-level cross section of $\mu\rightarrow\nu_{\mu}e^{-}\bar{\nu}_e$
  under the on-shell mass renormalization prescription, where $q_1\cdot q_2=q_2\cdot q_3=m_{\mu}^2/6-m_e^2/2$
  and $\sigma_0(\mu\rightarrow\nu_{\mu}e^{-}\bar{\nu}_e)$ is the tree-level cross section.}
\end{center}
\end{figure}
In Fig.15 the following data have been used: $e=0.3028$, $s_w^2=0.2312$, $m_W=80.42Gev$, $m_u=3Mev$, $m_c=1.25Gev$, $m_t=174.3Gev$, $m_d=6Mev$, $m_s=120Mev$, $m_b=4.2Gev$, $m_e=0.5110Mev$, $m_{\mu}=105.7Mev$, $m_{\tau}=1.777Gev$, $|V_{ud}|=0.975$, $|V_{us}|=0.223$, $|V_{ub}|=0.004$, $|V_{cd}|=0.222$, $|V_{cs}|=0.974$, $|V_{cb}|=0.040$, $|V_{td}|=0.009$, $|V_{ts}|=0.039$, and $|V_{tb}|=0.999$ \cite{c11}. Obviously the gauge dependence of $\sigma(\mu\rightarrow\nu_{\mu}e^{-}\bar{\nu}_e)$ induced by the on-shell mass renormalization prescription cannot be neglected at the two-loop level. On the other hand, from Eqs.(13,26) and Eq.(29) we find such gauge dependence doesn't appear in the result of the pole mass renormalization prescription.

\section{Conclusion}

Through calculating the singularities of W and Z's two-loop-level
transverse self energy we find the counterterms of W and Z's mass
and the sine of the weak mixing angle are gauge dependent under
the on-shell mass renormalization prescription. The gauge
dependencies of these counterterms lead to the cross section of
$\mu\rightarrow\nu_{\mu}e^{-}\bar{\nu}_e$ gauge dependent at
two-loop level. So the on-shell mass renormalization prescription
is a wrong mass renormalization prescription beyond one-loop
level.

On the other hand, all of the above gauge dependencies don't
appear in the results of the pole mass renormalization
prescription. So the pole mass renormalization prescription is the
only reasonable candidate for the mass renormalization
prescription at present. We should use the pole mass
renormalization prescription rather than the on-shell mass
renormalization prescription to calculate physical results beyond
one-loop level.

\vspace{5mm} {\bf \Large Acknowledgments} \vspace{2mm}

The author thanks Prof. Xiao-Yuan Li and Prof. Cai-dian Lu for their useful guidance.

\end{document}